\documentclass[longabstract,preprint,dvips,12pt]{aastex}

\usepackage[active]{srcltx}
\usepackage{natbib}
\usepackage{graphicx}
\usepackage[latin1]{inputenc}
\usepackage{amssymb}
\usepackage{amsmath}
\usepackage{xspace}
\usepackage{color}
\newcommand{\Ed}{E_{\rm d}}

\newcommand{\kt}{k_{\rm B}T}
\newcommand{\ktpar}{\kt_{\parallel}}
\newcommand{\ktperp}{\kt_{\hspace*{-0.15em}{\perp}}}
\newcommand{\NHp}{\ensuremath{{\rm NH}^{+}}\xspace}

\newcommand{\e}[1]{\ensuremath{\times 10^{#1}}}			

\title{Dissociative recombination measurements of NH$^+$ using an ion storage
ring
}

\author{O. Novotn\'y\altaffilmark{1}, 
	M.~Berg\altaffilmark{2},
	D.~Bing\altaffilmark{2},
	H.~Buhr\altaffilmark{2},
	W.~Geppert\altaffilmark{3},
	M.~Grieser\altaffilmark{2},
	F.~Grussie\altaffilmark{2},
	C.~Krantz\altaffilmark{2},
	M.B.~Mendes\altaffilmark{2}, 
	C.~Nordhorn\altaffilmark{2}, 
	R.~Repnow\altaffilmark{2},
	D.~Schwalm\altaffilmark{2,4},
	B.~Yang\altaffilmark{2,5}, 
	A.~Wolf\altaffilmark{2}, and
	D.~W.~Savin\altaffilmark{1}
	}

\email{oldrich.novotny@mpi-hd.mpg.de}

\altaffiltext{1}{Columbia Astrophysics Laboratory, Columbia University, New York, NY 10027, USA}
\altaffiltext{2}{Max Planck Institute for Nuclear Physics, 69117 Heidelberg,
Germany}
\altaffiltext{3}{Department of Physics, Stockholm University, AlbaNova, SE-106 91, Stockholm, Sweden}
\altaffiltext{4}{Faculty of Physics, Weizmann Institute of Science, Rehovot 76100, Israel}
\altaffiltext{5}{Institute of Modern Physics, Chinese Academy of Sciences,
Lanzhou 730000, China}


\begin{abstract}
We have investigated dissociative recombination (DR) of \NHp with
electrons using a merged beams configuration at the TSR heavy-ion storage ring
located at the Max Planck Institute for Nuclear Physics in Heidelberg, Germany.
We present our measured absolute merged beams recombination rate coefficient for
collision energies from 0 to 12~eV. From these data we have extracted a cross
section which we have transformed to a plasma rate coefficient for the
collisional plasma temperature range from $T_{\rm pl} = 10$ to $18000$~K. We
show that the \NHp DR rate
coefficient data in current astrochemical models are underestimated by up to a
factor of $\sim 9$. Our new data will result in predicted \NHp abundances 
lower than calculated by present models. This is in agreement with the
sensitivity limits of all observations attempting to detect \NHp in interstellar
clouds. 
\end{abstract}

\begin{document}
\doublespace
\maketitle

\section{Introduction}
\label{introduction}

Unraveling the molecular evolution of the Universe hinges, in part, on our
understanding of nitrogen and the associated chemistry. Nitrogen is the fifth
most cosmically abundant element \citep{Asplund:ARAA:2009}. Nitrogen-bearing
species account for $\sim 35\%$ of the molecules identified to date in space
\citep{AstrochymistMolecules}\ and can be found in variety of space environments
(e.g., \citealt{Wyckoff:ApJ:1991, Vuitton:ApJ:2006, Hily-blant:AA:2010,
Persson:AA:2010}). 

In the cold ISM, such as in molecular clouds, the most abundant nitrogen-bearing
species are expected to be N and N$_2$ \citep{Langer:ApJS:1989}. However, direct
observation of these is difficult. Atomic N does not have any low-lying,
fine-structure levels that can be populated at molecular cloud temperatures and
N$_2$ lacks a dipole moment, which is needed to provide reasonably strong
ro-vibrational transitions. Thus, observations of tracers such as NH, NH$_2$,
NH$_3$, or N$_2$H$^+$ must be used to infer the N and N$_2$ abundances through
chemical models. 

Neutral nitrogen hydrides are widely seen in the ISM. Ammonia (NH$_3$) was first
detected by \cite{Cheung:PRL:1968}.  Later NH$_2$ and NH were identified by
\cite{Dishoeck:ApJ:1993}\ and \cite{Meyer:ApJL:1991}, respectively. The observed
abundances, however, do not match predictions from astrochemical models. In
diffuse interstellar clouds, observed  abundance ratio for NH/NH$_3$ are $\sim
1.7$ and for NH$_3$/H  $\sim3.2\e{-9}$. These cannot be simultaneously explained
by existing chemical models. The models can fit either one of the observed
values but then the other predicted ratio is a factor of 10 off from the
observation \citep{Persson:AA:2010}. Similarly, in dark clouds the abundance
ratio for NH/NH$_3$ is underpredicted by more than an order of magnitude
\citep{Hily-blant:AA:2010}. These discrepancies possibly originate from using
incorrect reaction rate coefficients in the models. 

Key to understanding the nitrogen chemical network are the simple nitrogen
hydrides, both neutral and ionized
\citep{Pickles:ASS:1977,Almeida:AA:1982,Hily-blant:AA:2010,Persson:AA:2010,
Dislaire:AA:2012}. In the gas-phase, cold ISM these simple hydrides are
generated primarily within the ammonia formation pathway. The reaction chain
begins with \NHp production through
\begin{equation}
 {\rm N}^+ + {\rm H}_2 \rightarrow {\rm NH}^+ + {\rm H}. \label{eq:NHpform}
\end{equation}
Although this reaction is endoergic by $\sim 19$~meV and also dependent on both
the ortho versus para state of H$_2$ and on the fine-structure excitation of the
N$^+$, for all permutations it still proceeds even at temperatures as low as
10~K (\citealt{Dislaire:AA:2012, Zymak:ApJ:2013}; and references therein). It
has been also shown that 
\NHp can then react with H$_2$ to form yet more complex nitrogen hydride ions
via the series of reactions 
\begin{equation}
 {\rm NH}^+ \xrightarrow{{\rm H}_2} {\rm NH}_2^+ \xrightarrow{{\rm H}_2} {\rm
NH}_3^+ \xrightarrow{{\rm H}_2} {\rm NH}_4^+.  
\label{eq:ammonia2}
\end{equation}
Dissociative recombination (DR) of any of the last three ions with electrons
can form neutral nitrogen hydrides. Ammonia is formed by DR of NH$_4^+$
via the product channel
\begin{equation}
 {\rm e}^- + {\rm NH_4^+}  \rightarrow  {\rm NH_3 + H }.\label{eq:ammonia}
\end{equation}

Reactions (\ref{eq:ammonia2}) and (\ref{eq:ammonia}) indicate that \NHp is a key
molecule for gas-phase nitrogen chemistry; and several attempts have been
made to detect it in ISM
\citep{Snow:ASS:1979,Polehampton:MNRAS:2007,Hily-blant:AA:2010,
Persson:AA:2010,Persson:AA:2012}. None, however, have been successful. 
Our understanding of the role played by \NHp in forming more complex nitrogen
hydrides thus relies entirely on the accuracy of the relevant formation and
destruction rate coefficients needed for the astrochemical models.

An important destruction channel for \NHp is DR through
\begin{equation}
 {\rm e}^- + {\rm NH^+}  \rightarrow  {\rm N + H }.
  \label{eq:NHpDR}
\end{equation}
This reaction was previously investigated by \cite{Mcgowan:jPRL:1979}\ in a
single-pass merged beam experiment. In their measurement the ions did not have
time to relax and are expected to have been highly ro-vibrationally and
possibly also electronically excited.
Other experimental studies show that the DR rate coefficient can depend
significantly on the internal excitation of the ions (e.g.,
\citealt{Amitay:Science:1998}, \citealt{Mccall:PRA:2004}). Therefore, the
results of \citeauthor{Mcgowan:jPRL:1979}\ are unlikely to be applicable to
astrochemical models of the cold ISM where the ions are expected to be in their
vibrational ground state. Additionally, the lowest collision energies
investigated by \citeauthor{Mcgowan:jPRL:1979}\ correspond to a
collisional plasma
temperature of $\sim 100$~K; but modeling the cold ISM requires rate
coefficients down to temperatures of $\sim 10$~K. Extrapolating their results
down to these temperatures introduces additional uncertainty into the models.
Moreover, the original data of \citeauthor{Mcgowan:jPRL:1979}\ were later
corrected for an erroneous scaling used in the data analysis. The revised 
rate coefficient is smaller by a factor of $\sim2.5$
\citep{Mitchell:PR:1990,Mitchell:private:2012}. Unfortunately, databases used by
astrochemical models mix the original and the corrected rate coefficient values
(e.g., \citealt{Wakelam:ApJS:2012,Mcelroy_UMIST:AA:2013}).

In order to help improve our understanding of the nitrogen astrochemistry, we
have carried out DR measurements for \NHp. The rest of this paper is organized
as follows. The possible pathways for DR of \NHp are discussed in
Section~\ref{sec:DRpaths}. The experimental setup, measurement
method, and data analysis are described in Section \ref{exp_setup}. In Section
\ref{sec:res} we present the resulting merged beams DR rate
coefficient for \NHp, extract a DR cross section, and subsequently derive a 
Maxwellian plasma DR rate coefficient. We discuss our results and their
implications for astrochemistry in Section \ref{sec:discuss}. A summary is given
in Section \ref{sec:sum}. 
Additionally to the present study we have investigated the NH$^+$
DR at high collision energies. These complementary results were published by
\cite{Yang:JPB:2014}.

\section{DR pathways for NH$^+$} \label{sec:DRpaths}
In DR the incident electron is first captured into a doubly excited
state of the neutral molecule \citep{Bates:PR:1950,Bardsley:JPB:1968,
Larsson:book:2008}. If the potential surface of such a state is repulsive, the
neutral system formed can directly dissociate into neutral products.
Alternatively, if the electron is captured to a bound state it can
subsequently predissociate by coupling to a neutral repulsive state. These two
basic pathways are referred to as direct and indirect DR, respectively. In both
cases the overlap within the Frank-Condon region between the initial ionic state
and the intermediate neutral excited state defines the shape of DR cross section
versus electron-ion collision energy $E$. A more detailed description of the
various DR pathways is given in \cite{Larsson:book:2008}. Additionally, the
impact of molecular structure on the energy dependence of the DR cross section
has been described by \cite{Wolf:JPCS:2011} and \cite{Novotny:ApJ:2013}. 
In this section we discuss the molecular structure of both \NHp and NH
and review some of the expected energy dependence for DR of NH$^+$.

The structure of NH and \NHp has been studied both spectroscopically and
theoretically (e.g., \citealt{Colin:CJP:1968, Wilson:Nature:1978,
Goldfield:JCP:1987, Kawaguchi:JCP:1988, Colin:JMS:1989,
Clement:JCP:1992,Palmieri:JPC:1996, Amero:IJQC:2005, Owono:JCP:2007,
Hubers:JCP:2009, Beloy:PRA:2011}). To the best of our knowledge, however, no
potential energy curves have been published for the repulsive doubly excited NH
states. Nevertheless, these states are expected to form various Rydberg series
with each series converging to a single repulsive ionic core state as the
principle quantum number $n$ of the captured electron goes to $\infty$. The two
\NHp repulsive ionic core states relevant for this study are the $2\,^2\Pi$ and
$2\,^4\Sigma^-$, with vertical excitation energies in the Franck-Condon region
of $\sim 7.7$~eV and $\sim 10.5$~eV, respectively (\citealt{Amero:IJQC:2005};
a summary on known NH and NH$^+$ of potential curves relevant for NH$^+$ DR can
be found in \citealt{Yang:JPB:2014}).
The Rydberg series of neutral doubly excited repulsive states converging to
these thresholds can lead to direct DR. The non-resonant nature of the
transition between the initial ionic state and the repulsive neutral state
typically results in broad features in the DR energy spectrum.

Indirect DR proceeds via neutral bound states. These can also be
grouped into different Rydberg series, each converging to a particular ionic
core level. The excitation energy of each ionic core forms an
upper limit for a given series of indirect DR resonances. Individual resonances
are often experimentally unresolvable. However, the series limits may appear
more clearly in the measured cross section versus collision energy. If the
neutral Rydberg states contribute significantly to DR, then the DR signal will
decrease as the collision energy scans over the ionic threshold. On the other
hand, the neutral Rydberg states may also lead to additional autoionization
channels or can be an initial step for various other reactions, such as
dissociative excitation or ion-pair formation, which compete with DR. 
This can lead to a reduction in DR from other channels (e.g., direct DR) through
the given neutral Rydberg state. In an extreme case such DR reduction through
Rydberg states may appear as a DR increases above the ionic limit. From the
available data on the \NHp structure we cannot predict which of these cases will
dominate for particular Rydberg series.

For indirect DR there are a number of different \NHp ionic core thresholds which
can be relevant to DR at the lowest collision energies. These thresholds include
the $\rm X\,^2\Pi$ excited rotational $J$ levels, fine structure excitation to
the $\rm X\,^2\Pi_{3/2}$ level by $\sim13$~meV, and the $\sim42$~meV excitation
to the $\rm a\,^4\Sigma^-$ electronic state \citep{Hubers:JCP:2009}. The
vibrational $v$ levels of the \NHp $\rm X\,^2\Pi$ and $\rm a\,^4\Sigma^-$ states
can also play a role.  The level spacing for these two vibrational series is
approximately 0.38~eV and 0.33~eV, respectively \citep{Amero:IJQC:2005}. The
next higher lying \NHp bound electronic states are the $\rm A\,^2\Sigma^-$, $\rm
B\,^2\Delta$, and $\rm C\,^2\Sigma^+$. The corresponding thresholds for indirect
DR from ground-state \NHp to the lowest vibrational levels of these electronic
states may appear at $E\approx2.7$~eV, 2.8~eV, and 4.3~eV, respectively
\citep{Colin:CJP:1968}. Note that neutral resonances corresponding to 
electron capture into Rydberg levels attached to these electronic state
thresholds extend several electron volts below these thresholds and can hence be
important over a large range of electron collision energies.

DR of \NHp is exothermic. The amount of available energy constrains both the
maximum internal excitation of the products and also the number of possible
dissociation pathways. The DR cross section may increase above the threshold for
forming excited atomic products as a result of the opening up of new
dissociation channels. To evaluate the exothermicity for DR of \NHp we briefly
review the relevant thermochemical data. The \NHp ground state ${\rm
X}\,^2\Pi_{1/2}(v=0, J=1/2)$ lies $3.524\pm0.003$~eV below the $\rm N + H^+$
dissociation limit \citep{Marquette:JCP:1988}. The ground state $\rm N+H$
products lie energetically lower by the hydrogen ionization energy of 13.598~eV
\citep{NISTWebBookWhole}. From these values we obtain an \NHp DR exothermicity
at $E=0$~eV of $10.074\pm0.003$~eV. This energy is shared between the product
electronic excitation and the kinetic energy released (KER) of the DR products.
At $E=0$~eV the exothermicity is insufficient to excite the H atom and it is
produced in the $\rm ^2S_{1/2}$ ground state. On the other hand, for nitrogen
there are three electronic terms available, namely the $\rm ^4S^o$ ground term
and the excited $\rm ^2D^o$ and $\rm ^2P^o$ terms, with KERs of 10.074~eV,
7.690~eV, and 6.498~eV, respectively. Additional channels open at higher
collision energies. The channels most important for discussing our \NHp results
open at collision energies below $\sim 1.6$~eV. The channels relevant for this
energy range are listed in Table~\ref{tab:channels}. Channels which play a role
at higher energies can readily be obtained from the atomic data of
\cite{AtomicSpectraDatabase}. It should be emphasized that all exothermicities
are given for \NHp being in its ground state. Any internal excitation of the
\NHp results an increase in the exothermicity of the reaction. This internal
excitation shifts the collision energy thresholds for opening the product
channels with indices ${\rm ID}\geq4$ (see Table~\ref{tab:channels}) to
correspondingly  lower values  (less negative KER in Table~\ref{tab:channels}).
Experimentally we expect internal excitation to result in a smoothing with $E$
of the threshold behavior.  This is a consequence of the stored ions coming into
thermal equilibrium with the ambient 300~K temperature of TSR, resulting in a
range of rotational levels being populated.

There are other electron-induced processes that could compete with DR and thus
reduce the DR signal. For the case of \NHp these competing processes are all
endothermic. One of these is ion-pair formation which can yield 
N$^+$+H$^-$ at energy above $\sim3.7$~eV. The complementary channel N$^-$+H$^+$
is not expected to exist due to instability of N$^-$ \citep{NISTWebBookWhole}.
Another electron-driven process is dissociative excitation (DE) forming N +
H$^+$ or N$^+$ + H.  These reactions are endothermic by $\sim3.5$~eV and
$\sim4.5$~eV, respectively.

\section{Experimental}
\label{exp_setup}
\subsection{Setup}\label{sec:setup}

Data were collected in July 2010 using the TSR storage ring located at the Max
Planck
Institute for Nuclear Physics (MPIK) in Heidelberg, Germany. Details on various
aspects
of this merged-beams experimental setup have been described in more detail by
\cite{Amitay:PRA:1996}, \cite{Wolf1999}, and \cite{Novotny:ApJ:2013}. Here we
discuss only those aspects specific to this study.

To generate the \NHp ion beam, we first produced NH$_3^-$ in a
cesium sputtering source with a molybdenum target and ammonia as the parent gas.
The negative ions were
accelerated and then directed through a nitrogen gas stripping curtain. The
resulting \NHp ions were further accelerated to a final energy of $E_{\rm
i}\approx 6.2$~MeV and injected into the storage ring. Typical stored ion
currents were $\sim 0.3$~nA during data acquisition.

The stored ion beam was merged with two nearly mono-energetic electron 
beams, dubbed the Target \citep{Sprenger:NIMPRA:2004} and the Cooler
\citep{Steck:NIMA:1990}. Starting after ion injection at time $t=0$,
both electron beams were velocity matched to the ions  until $t = 6$~s. During
this time
elastic collisions of the ions with the low energy spread electron
beams transferred energy from the recirculating ions to the single pass
electrons, a process known as electron cooling \citep{Poth:PR:1990}. This
reduced the energy spread of the ions and resulted in a narrow ion beam diameter
($<1$~mm). Data were collected from $t = 6$~s to $t = 16.5$~s. This was
then followed by next ion injection and the cycle repeated.

The energy spread  of each electron beam is critical for the energy resolution
of the experiment. The energy distribution is parametrized by the effective
temperatures $T_\perp$ and $T_{||}$, respectively perpendicular and parallel to
the bulk electron beam velocity. The Target electron beam was produced using a
photocathode in transmission mode \citep{Orlov:NIMA:2004, Orlov:JAP:2009}\ and
expanded adiabatically by a factor of 20. For these conditions we expect
$\ktperp^{\rm T}=1.65\pm0.35$~meV and $\ktpar^{\rm T}= 25^{+45}_{-5}~\mu$eV
(\citealt{Novotny:ApJ:2013}), where $k_{\rm B}$ is the Boltzmann constant. Here
and throughout all uncertainties are quoted at an estimated $1\sigma$
statistical confidence level. The Cooler uses a thermionic emission cathode and
expansion factor of 9.6. The corresponding electron beam temperatures are
$\ktpar^{\rm{C}}\approx180~\mu$eV and $\ktperp^{\rm{C}}\approx13.5$~meV
\citep{Lestinsky:prl:2008}.

After the cooling phase, the Target electron beam was used as a probe to measure
DR over a range of electron-ion center of mass collision energies.  This was
accomplished by varying the laboratory energy of the Target. While this was
happening, the Cooler beam remained velocity matched with the ions at all times.
In this way the ion beam was continuously cooled and did not expand in size or
shift in energy as the Target beam was detuned. The Target and Cooler were
operated with typical electron beam densities of $n_{\rm e}^{\rm
T}\approx2.7\e{6}$~cm$^{-3}$ and $n_{\rm e}^{\rm C}\approx2.5\e{7}$~cm$^{-3}$,
respectively.
Neutral DR products generated in the Target were not
deflected by the first dipole magnet downstream of the Target and continued
ballistically until they hit a detector. The detected signals provided the
event countrates.

\subsection{NH$^+$ internal excitation}
\label{sec:NHpexc}
Ions produced by stripping are expected initially to possess
electronic, vibrational, rotational, and fine-structure excitation. We estimate
that radiative relaxation removes most of the \NHp internal excitation during
the 6~s of electron cooling, resulting in a parent ion population far closer to
astrophysical conditions than that which was achievable in the single pass
results of \cite{Mcgowan:jPRL:1979}. Below the first \NHp dissociation limit lie
the $\rm a\,^4\Sigma^-$, $\rm A\,^2\Sigma^-$, and $\rm B\,^2\Delta$
excited electronic states.
The measurements of \cite{Brzozowski:PS:1974}\ yielded decay times for the $\rm
A\,^2\Sigma^-$ and $\rm B\,^2\Delta$ states of $\sim 1~\mu$s. Thus, these states
are expected to have fully decayed before the onset of data acquisition. To the
best of our knowledge, there have been no such investigations into the radiative
lifetime of the $\rm a\,^4\Sigma^-$ symmetry. Moreover, even if this state
relaxes quickly into thermal equilibrium with the $\sim 300$~K vacuum chamber,
the expected Boltzmann distribution indicates that $\sim 15\%$ of the \NHp will
be in the $\rm a\ ^4\Sigma^-$ state. 

We are not aware of any theoretical or experimental investigations into
radiative relaxation of the \NHp ro-vibrational levels and have therefore
calculated them ourselves. We predict that all of the ions relax to their $v=0$
level and that the rotational excitation comes into equilibrium with the
$\sim300$~K black-body radiation of the vacuum chamber during the initial
electron cooling phase. To model this we have calculated the ro-vibrational
radiative lifetimes of $\rm X\,^2\Pi_{1/2}$ for levels ranging from $v=1$ to $v
= 5$ and from $J=3/2$ to $J = 19/2$. Higher levels are calculated to decay so
rapidly that they are irrelevant for the model. Our approach uses a method
similar to that of \cite{Amitay:PRA:1994}. The dipole moment for \NHp was taken
from \cite{Cheng:PRA:2007}. With this as a guide, we have generated a radiative
relaxation model using the calculated spontaneous radiative decay lifetimes
while also accounting for stimulated emission and absorption by the 300~K
black-body radiation. For the initial rotational excitation we have taken a
Boltzmann distribution at temperature of 8000~K. This is approximately the
excitation temperature derived in a similar DR experiment on CF$^+$
\citep{Novotny:JPCS:2009}. After the initial 6~s of ion storage, the model
predicts all vibrational levels decayed to $v=0$. The remaining average
rotational excitation energy exceeds the 300~K equilibrium by only 12\%. The
excitation energy averaged over the ion population during the measurement
window, from $t = 6.0 - 16.5$~s, exceeds room temperature excitation by only
4.6\%. This predicted level of excitation might be overestimated due to the
omission of spin-orbit coupling and coupling to the $\rm a\,^4\Sigma^-$ state,
both of which may result in extra decay pathways. The model also does not take
into account additional acceleration of the rotational cooling from
super-elastic ion collisions with electrons (e.g., \citealt{Shafir:PRL:2009}).

The final excitation to consider is the fine structure splitting $\rm
X\,^2\Pi_{1/2-3/2}$. This is not expected to relax during ion storage. We are
unaware of any published lifetime estimates for this transition. Moreover, the
excitation energy of $\sim 13$~meV is well within the energies accessible by the
$\sim 300$~K ambient radiation. Taking into account the Boltzmann distribution
at 300~K for all the rotational levels in the $\rm X\,^2\Pi_{1/2}$  and
$\rm X\,^2\Pi_{3/2}$ fine-structure levels \citep{Kawaguchi:JCP:1988}, we
predict that the two fine-structure groups are statistically populated by $\sim
67\%$ and $\sim 33\%$, respectively.

\subsection{Measurements}\label{sec:meas}
During the data acquisition phase 
of each injection cycle, the Target electron beam energy was stepped over a
repeating set of three different energies. 
For the first step, cooling, the Target electrons were velocity matched to the
ions at an electron energy of $E_{\rm cool}=226.74$~eV in the
laboratory frame. Although the Cooler beam continuously cooled the ions, the
much narrower energy spread of the Target beam provided even stronger cooling
during this step. This helped to maintain a stable phase-space spread
of the ions during storage. 

In the next step, measurement, the Target beam was set 
to a mean energy of $E_{\rm meas}$ in the laboratory frame. The center of mass
collision energy, the so-called detuning energy $\Ed$, can be calculated
non-relativistically from the mean electron and ion beam velocities, or from
corresponding laboratory frame ion beam energies as
\begin{equation}
E_{\rm d} = \left(\sqrt{E_{\rm meas}} - \sqrt{E_{\rm cool}}\right)^2. 
\label{eq:ed}
\end{equation}
The count rate at the detector during this step was used to obtain the merged
beams DR rate coefficient at the calculated $\Ed$. The measurement energy, and
hence detuning energy, was changed for each injection cycle.

In the last step, reference, the Target beam energy was set 
to a detuning energy $\Ed^{\rm ref}\approx 29$~meV. This reference energy was
identical for all injection cycles. The detector count rate during this step was
used to monitor the ion beam intensity. 

The laboratory frame electron beam energies for each step were obtained from the
Target cathode voltage which was then corrected for space charge effects
\citep{Kilgus:PRA:1992}. The power supply providing the cathode voltage
requires a short time to equilibrate. Thus to ensure stable conditions
during data acquisition at each step we add 5~ms settling time between setting
the new cathode voltage and the start of the corresponding step.
Excluding this settling time, the dwell times at the cooling, measurement, and
reference steps were 40~ms, 30~ms, and 30~ms, respectively.

The count rate of the DR events was measured using a $10\times10$~cm$^2$ energy
sensitive Si surface-barrier detector located $\sim12$~m downstream of the
Target. For each DR event the N and H fragments arrive at the detector with a
time difference of only a few nanoseconds. Such a small delay cannot be
distinguished by the detector and hence only a single pulse is detected for each
DR event, representing the total kinetic energy of the fragments.

Typically both fragments reached the detector and the measured
total kinetic energy corresponds to the original ion beam energy. However, DR
of \NHp can release several eV in kinetic energy (see Table~\ref{tab:channels}).
Such a kick
in the direction perpendicular to the parent ion beam can deflect the light H
fragments sufficiently so that they miss the active area of the detector. On the
other hand, the kinetic energy released has little effect on the heavier N
fragments. They remain confined to a narrow cone which is fully enclosed by the
detector. The DR signal is therefore a sum of detections of both fragments and
detections of only the N fragment. 

A number of background processes can mimic our signal and must be account for.
For example, electron-ion collisions can result in DE producing N and H$^+$
fragments. The threshold for this DE channel is $\sim 3.5$~eV. Thus for $\Ed$
below this, to determine the DR signal we can use events with kinetic energies
corresponding to capturing both  fragments (N+H) and also those detecting only
the N fragments. At higher energies, however, DE events could contaminate the DR
data. Hence, above $\sim 3.5$~eV we used only events where both N and H reached
the detector. These data need then to be adjusted to account for those DR events
where the H fragment does not hit the detector due to the detector size. This
correction was measured independently using a position sensitive detector
\citep{Buhr:PRA:2010, Yang:JCP:2014}. The resulting rescaling of the DR signal
was less than 7\% over the energy range studied.

The DR signal was also contaminated by reaction products from various collisions
of the ions with the residual gas. Dissociative charge transfer produces neutral
N and H fragments. Collision induced dissociation can produce either N and H$^+$
fragments or N$^+$ and H fragments. In all of these cases the resulting N+H and
N events on the neutral-fragment detector can be misinterpreted as DR and need
to be accounted for.  To do this we used the fact that the rate of these
background events, $R_{\rm BG}$, scales with the ion beam intensity and with the
residual gas pressure, but -- in contrast to DR -- it does not depend on the
electron density. Thus any signal that does not depend on the electron beam
density can be used as a proxy for the total DR background rate at the
neutral-fragment detector induced by residual gas collisions. Here we use the
signal from N$^+$ fragments produced in collision induced dissociation to
monitor the relative ion beam intensity. For this we employ a detector located
in the first dipole magnet downstream the Target \citep{Lestinsky2007}. This
detector was positioned such that only N$^+$ fragments originating in the Target
were deflected onto its active area. We used only the data from the reference
step which is at an energy below the threshold for DE forming N$^+$ + H. A
scaling factor $\xi$ is needed to match the N$^+$ count rate, $R_{{\rm N}^+}$,
to the background count rate $R_{\rm BG}$, so that $R_{\rm BG} = \xi R_{{\rm
N}^+}$. 

We determined $\xi$ independently using a special measurement scheme.
After ion injection and the cooling phase, we switched off the Target electron
beam. The ion beam was still cooled by the Cooler. With the Target off, the
count rate at the neutral fragment detector originated only from residual gas
induced reactions and was thus equal to $R_{\rm BG}$. Simultaneously we acquired
also the $R_{{\rm N}^+}$ signal. The scaling factor $\xi$ was simply obtained
by comparing $R_{\rm BG}$ to $R_{{\rm N}^+}$ at the same storage times.

We determined the measured merged beams DR rate coefficient $\alpha_{\rm
mb}(\Ed)$ by normalizing the recorded DR signal by the electron density and ion
number (e.g., \citealt{Amitay:PRA:1996}). The electron beam density can be
determined from the measured electron beam current, energy, and
geometry \citep{Sprenger:NIMPRA:2004}. The number of interacting ions can be
calculated from the ion beam current. However, the typical stored \NHp beam
current was only a few nA. This is several orders of magnitude lower than what
can be measured directly using the available DC current transformer
\citep{Unser:IEEETNS:1981}. Hence we first determined a relative DR rate
coefficient versus $\Ed$ using the $R_{{\rm N}^+}$ signal as a relative proxy
for the ion beam current. We then scaled the whole curve to match the absolute
rate coefficient value determined at $\Ed=0$~eV using an independent measurement
technique based on comparing ion beam decay rates with and without the electron
beam present in the interaction zone \citep{Novotny:APJ:2012}. 

\subsection{Generating a DR Cross Section and a Plasma DR Recombination Rate
Coefficient}
\label{sec:PlasmaRateCalc}
In order to enable researchers to use  our storage ring data in astrochemical
models we have converted the measured merged beam DR rate coefficient into a
plasma DR rate coefficient.
The transformation is a two step process. First the experimental collision
energy distribution is deconvolved from $\alpha_{\rm mb}(\Ed)$ to obtain a cross
section $\sigma(E)$ as a function of collision energy $E$. In the second step
the cross section is convolved with a Maxwell-Boltzmann distribution to
generate a plasma rate coefficient $\alpha_{\rm pl}(T_{\rm pl})$ at the required
collisional plasma temperature $T_{\rm pl}$. Here $T_{\rm pl}$ reflects
the collisional velocity spread in a plasma at given
temperature, but not the internal excitation of the ions.
We have recently developed a novel
method for converting $\alpha_{\rm mb}(\Ed)$ to $\sigma(E)$ and subsequently
into $\alpha_{\rm pl}(T_{\rm pl})$. The procedure allows also for propagating
statistical and systematic uncertainties. The method is described in detail by
\cite{Novotny:ApJ:2013}.

\section{Results}
\label{sec:res}
\subsection{Merged Beams Recombination Rate Coefficient}
\label{sec:resrate}

Figures~\ref{fig:rateexpA} and \ref{fig:rateexpB} present our merged beams
rate coefficient for DR of \NHp. In Figure~\ref{fig:rateexpA} we display the
data on a log-log scale for the measured detuning energies $\Ed=15~\mu$eV to
12~eV. These data are also given in tabular form in Table~\ref{tab:data:MB}. In
Figure~\ref{fig:rateexpB} we zoom in on the most pronounced rate coefficient
structures in the detuning energy range from $\Ed=40$~meV to 5~eV and plot the
data on a lin-log scale.  

The systematic uncertainty in the measured $\alpha_{\rm mb}$ has
several important components. The uncertainties from the ion beam storage life
time measurements and from the electron density dominate the 8.4\% uncertainty
on the absolute scaling of $\alpha_{\rm mb}$. This value is independent of
$\Ed$. Another source of error is the background subtraction factor $\xi$. At
energies $\Ed<0.01$~eV, the corresponding uncertainty propagated to $\alpha_{\rm
mb}$ amounts to $\lesssim 1$\%. As the collision energy increases, the
uncertainties are $\sim 5$\%, $\sim 10$\%, $\sim 50$\%, and $\sim 7$\% at
$\Ed=0.1$~eV, $1.5$~eV, $3.5$~eV, and $10$~eV, respectively. The energy
dependence of the uncertainty is due to the decrease in DR up to  3.5 eV
followed by an increase going to  10 eV, while the background stays constant. At
$\Ed>3.5$~eV an additional error originates from correcting for geometrical
detector losses. We estimate this uncertainty to be less than 1\%. The total
systematic error is then $\lesssim 8.5$\%, $\sim 10$\%, $\sim 13$\%, $\sim
50$\%, and $\sim 11$\% at $\Ed\lesssim0.01$~eV as well as $\Ed=0.1$~eV,
$1.5$~eV, $3.5$~eV, and $10$~eV, respectively.

For each data point in $\alpha_{\rm mb}$ the statistical uncertainty is given by
the counting statistics from the number of signal and background counts. This
error amounts to $\lesssim 4$\%, $\sim 10$\%, $\sim 50$\%, and $\sim 8$\%
at $\Ed\lesssim1$~eV as well as  $\Ed=1.5$~eV, $3.5$~eV, and $10$~eV,
respectively.

\subsection{Recombination Cross Section}\label{sec:rescrosssec}

We have converted the experimental DR rate coefficient $\alpha_{\rm
mb}$ to a cross section $\sigma$. The result is plotted in
Figure~\ref{fig:res_crosssec}.  These data are also given in tabular form in
Table~\ref{tab:data:CS}. The lower edge of the first energy bin is
set to $E=0$ and is not displayed on the logarithmic energy scale of the
plot.

For the conversion of $\alpha_{\rm mb}$ to $\sigma$ we followed the procedure
developed by \cite{Novotny:ApJ:2013}. The method involves fitting  $\alpha_{\rm
mb}$ with a model rate coefficient. The fitting yielded a minimum
chi-squared of $\chi^2/N_{\rm
DF} = 1.49$, where $N_{\rm DF} = 46$ is the number of degrees of freedom in the
fit. The conversion procedure allows for propagating the uncertainties of
$\alpha_{\rm mb}$ to the cross section. The systematic errors from the absolute
scaling, the background subtraction, and the correction on the geometrical
detector efficiency propagate all nearly directly, i.e., relative errors in
$\sigma$ at $E$ are close to those in $\alpha_{\rm mb}$ at $\Ed = E$. The
statistical errors also propagate similarly, except for  energies
$\Ed\lesssim\ktpar^{\rm T}$ where the statistical uncertainty increases up to
30\%.

The uncertainties in the electron beam parameters, primarily in $\ktpar^{\rm T}$
and $\ktperp^{\rm T}$, also affect the cross section error. We calculate this
error by modeling the experimental electron-ion collision energy distribution.
The corresponding uncertainties in $\sigma$ are displayed as gray bars in
Figure~\ref{fig:res_crosssec}. The largest sensitivity can be seen at collision
energies $\Ed\lesssim\ktpar^{\rm T}$, where the collision energy spread is
comparable to the mean collision energy. Some enhanced uncertainties appear also
at $E\approx0.1$~eV. This is due to the cross section bins being comparable in
width to the energy spread at these energies.

\subsection{Plasma Recombination Rate Coefficient}\label{sec:resplasma}
\begin{figure}
    \plotone{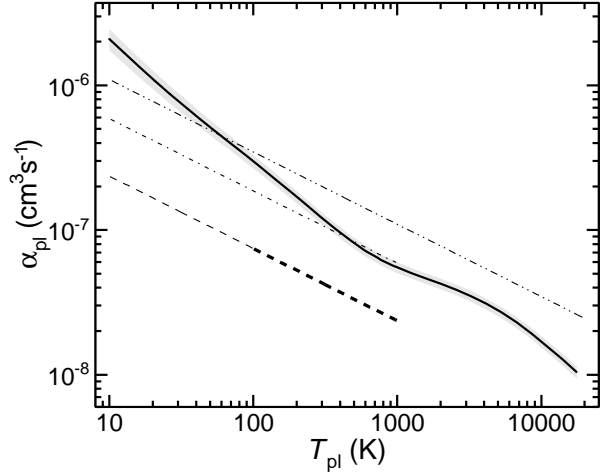}
    \caption{\doublespace Experimentally derived DR plasma rate coefficient
for \NHp plotted versus the collisional plasma temperature (full black line).
The gray band marks the total systematic uncertainty originating from the error
on the absolute scaling, background subtraction, detector geometrical efficiency
correction, and uncertainties in $\ktperp^{\rm T}$ and $\ktpar^{\rm T}$. The
statistical errors propagated from $\alpha_{\rm mb}$ are smaller than 1\% at all
temperatures and thus indistinguishable in the plot. The thick dashed line plots
the previous DR \NHp results of \cite{Mcgowan:jPRL:1979}, corrected for an
erroneous form factor \citep{Mitchell:PR:1990,Mitchell:private:2012}, in the
approximate temperature range of validity. The thin dashed line extrapolates
these data down to $T_{\rm pl}=10$~K as it is done in current astrochemical
models. The thin dot-dashed line plots the same curve but without the correction
for the erroneous form factor. These uncorrected data are used in some of the
astrochemical databases. The dot-dot-dashed
line plots $\alpha_{\rm pl}^{\rm di}$ which represents a ``typical'' rate
coefficient for diatomic ions.
	\label{fig:rateplasmac2}
	}
\end{figure}

We have converted $\sigma$ to a plasma rate coefficient using the procedures
described in Section \ref{sec:PlasmaRateCalc}. The resulting $\alpha_{\rm
pl}(T_{\rm pl})$ is plotted in Figure~\ref{fig:rateplasmac2} for a plasma
temperature range $T_{\rm pl}=10-18000$~K. Note that $T_{\rm pl}$ represents
only the spread in the collision energies between the electrons and ions. As
discussed in Sec.~\ref{sec:NHpexc}, the internal excitation of the studied ions
is expected to be $\sim 300$~K. This is exclusively due to rotational
excitation, while electronic and vibrational NH$^+$ excitation is negligible.

The statistical uncertainties propagate to $\alpha_{\rm pl}$ at very low
levels, amounting to less than 1\% at all temperatures. This is due to
the integrative nature of the transformation from $\sigma$ to $\alpha_{\rm pl}$.
{\color{red} Three systematic errors are relevant in the given temperature range
for the plasma rate coefficient. First, the absolute scaling error of
$\alpha_{\rm mb}$ propagates directly to $\alpha_{\rm pl}$ as an 8.4\% relative
error, which describes a single scaling of the complete $\alpha_{\rm pl}(T_{\rm
pl})$ curve. Second, the background subtraction uncertainty generates an
absolute error in $\alpha_{\rm pl}$ of $\pm7\e{-10}~\rm cm^3\,s^{-1}$,
independent of $T_{\rm pl}$. Lastly, the uncertainties of the electron beam
parameters propagate to $\alpha_{\rm pl}$ as an absolute error which can be well
approximated by $\pm 6.1\e{-6}\, (T_{\rm pl}/{\rm K})^{-1.3}~\rm cm^3\,s^{-1}$.
The behavior of the various systematic errors needs to be taken into account
when propagating the resulting rate coefficient uncertainties.  For chemical
models involving a range of temperatures, the three systematic errors must be
applied to all $\alpha_{\rm pl}(T_{\rm pl})$ points in a correlated way
according to the description given. For models at a single fixed
temperature, their appropriate absolute values can be added in quadrature.  This
results in a systematic error of $\sim16$\% at $T_{\rm pl}=10$~K, $\sim10$\% at
100~K, $\sim8$\% at 1000~K, $\sim 9$\% at 10000~K, and $\sim 9$\% at 18000~K.
}

To parametrize the results for use in astrochemical modeling, we have fit the
data with an analytical formula. We tried to fit the data with both the
two- and three-parameter functions commonly used to describe DR
plasma rate coefficients in astrochemical models and databases (e.g.,
\citealt{Florescu:PR:2006}, \citealt{kida-base}) but were unable to fit our
measured plasma rate coefficient over the entire temperature range with a
precision of any better than 40\%. Therefore we used the form proposed by
\cite{Novotny:ApJ:2013}, namely
\begin{equation}
\alpha_{\rm pl}^{\rm fit}(T_{\rm pl}) =
A\,\left(\frac{300 ~\rm K}{T_{\rm pl}}\right)^n + B ,
\label{eq:plasmafitnew}
\end{equation}
where
\begin{equation}
B = T_{\rm pl}^{-3/2}\sum_{i=1}^4 c_i \exp(-T_i/T_{\rm pl}).
\label{eq:B}
\end{equation}
We fit Equation \ref{eq:plasmafitnew} to our data over the full
temperature range.
The resulting parameters are given in Table~\ref{tab:plasmares}.
The deviations of $\alpha_{\rm pl}^{\rm fit}$ from the data are less than
1\% over the reported temperature range.

\section{Discussion}
\label{sec:discuss}

\subsection{Experimental DR rate coefficient and cross section}

\begin{figure}
\epsscale{0.5}
    \plotone{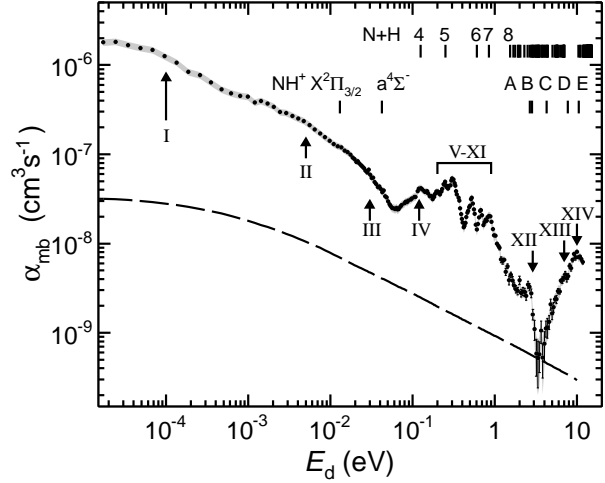}
    \caption{\doublespace
Experimental merged beams rate coefficient $\alpha_{\rm mb}$ for DR of \NHp. The
filled circles show the data and the error bars indicate the 1$\sigma$
statistical confidence level. The total 1$\sigma$ systematic error is marked in
gray. The long dashed line illustrates the shape of a merged beams rate
coefficient expected for a direct DR process, i.e., for a cross section
$\sigma(E)\propto E^{-1}$, and is arbitrarily scaled on the vertical axis. The
uppermost series of vertical lines marks the openings of various \NHp DR product
excitation channels. The energy thresholds are given for \NHp in its ground
state. The channel ID numbers are given in Table~\ref{tab:channels}. The
vertical lines in the second row  mark the excitation energies for various \NHp
states. The $\rm X\,^2\Pi_{3/2}$ and $\rm a\,^4\Sigma^-$ levels are labeled
directly in the figure. The labels A, B, C, D, and E correspond to $\rm
A\,^2\Sigma^-$, $\rm B\,^2\Delta$, $\rm C\,^2\Sigma^+$, $2\,^2\Pi$ and
$2\,^4\Sigma^-$ states, respectively. Roman numbers label various features in
the DR spectrum which are discussed in the text.
	\label{fig:rateexpA}
	}
\end{figure}

\begin{figure}
    \plotone{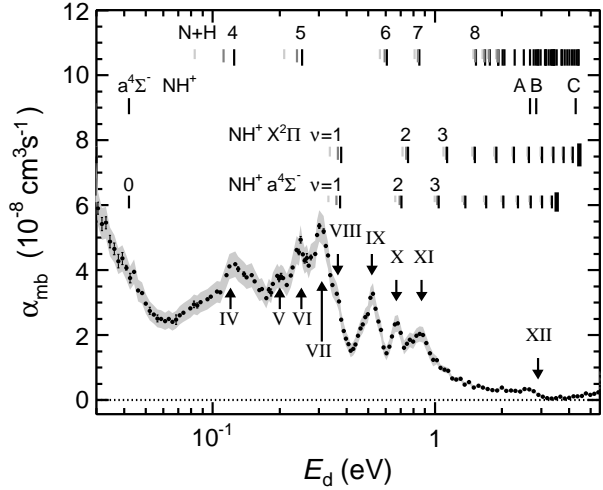}
    \caption{\doublespace
Same as Figure \ref{fig:rateexpA} but plotted for $\Ed \approx 0.04-5.0$~eV on
a lin-log scale. Additionally we have extended the number
of plotted energy thresholds. The first row of black vertical lines marks
openings of \NHp DR product excitation channels for \NHp in its ground state.
The shorter dark-gray lines then assume \NHp initially in the $\rm
X\,^2\Pi_{3/2}$ excited 
fine structure level. The yet shorter light-gray lines assume \NHp initially in
the $\rm a\,^4\Sigma^-(v=0)$ state. The vertical lines in the second row mark
the excitation energies for various NH$^+$ states. The third and fourth series
of vertical lines label vibrational thresholds for \NHp $\rm X\,^2\Pi$ and $\rm
a\,^4\Sigma^-$, respectively. The black, dark-gray, and light-gray encoding is
identical to the first row of vertical lines. Thick lines at the end of each
vibrational series mark the respective \NHp dissociation limits. 
	\label{fig:rateexpB}
	}
\end{figure}

\begin{figure}
    \plotone{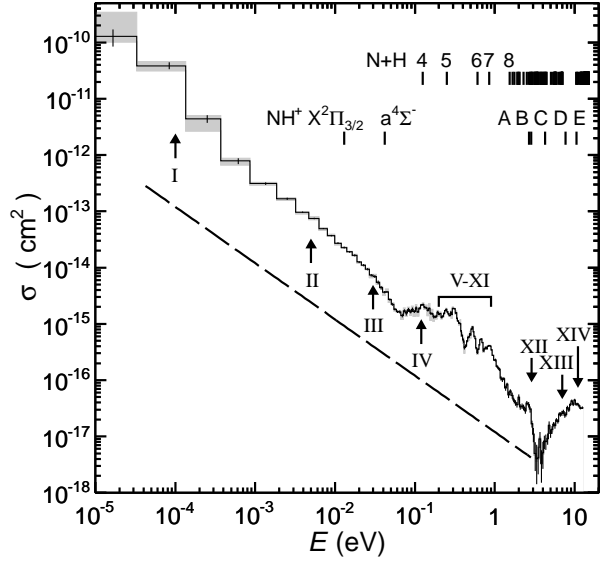}
    \caption{
	\label{fig:res_crosssec}
\doublespace 
The DR cross section for \NHp is shown by the solid  line histogram. The
lower edge of the left-most energy bin at $E=0$ eV is not shown here. The
vertical error bars describe the statistical uncertainty propagated from
$\alpha_{\rm mb}$ and potential numerical instabilities from the conversion
procedure (for more details see \citealt{Novotny:ApJ:2013}). The gray error
bands show the error originating from uncertainties in $\ktperp^{\rm T}$ and
$\ktpar^{\rm T}$. The long-dashed line illustrates the shape of the merged beams
rate coefficient expected for a direct DR process, i.e., $\sigma(E)\propto
E^{-1}$. The curve is arbitrarily scaled on the vertical axis. The vertical
lines and labels are identical to Figure~\ref{fig:rateexpA}.}
\end{figure}

The  evolution of the \NHp DR rate coefficient with collision energy displays
diverse features which can be used to probe the underlying quantum nature of the
process. 
For clarity we use Roman numerals to label the features in
Figures~\ref{fig:rateexpA}, \ref{fig:rateexpB}, and \ref{fig:res_crosssec} which
we discuss below. We also plot the thresholds for the opening of various DR
product excitation channels. Additionally we include in the figures the energies
of various \NHp states, which are upper limits for Rydberg series of NH states.
As we have discussed in Section~\ref{sec:DRpaths}, these neutral states are
potentially involved in both the direct and indirect DR processes and their
energies may be reflected in the DR energy spectrum. 

The overall shapes of $\alpha_{\rm mb}$ and $\sigma$ are dominated by a
decreasing magnitude with increasing energy, roughly following the $\sigma
\propto E^{-1}$ expected for direct DR process \citep{Florescu:PR:2006,
Larsson:book:2008}. Numerous structures appear on top of this global
shape. Three broad features (labeled I, II, and III) are visible at energies
below $\sim0.05$~eV. The features at these low collision energies can be caused
by electron capture to rotationally excited NH Rydberg states converging to the
\NHp $\rm X\,^2\Pi_{1/2}$, $\rm X\,^2\Pi_{3/2}$, and $\rm a\,^4\Sigma^-$ ionic
cores. The decreasing strength of these features with increasing collision
energy can be attributed to the opening up of additional autoionization
channels, though
individual resonances cannot be distinguished due to the experimental energy
resolution. Similar features, associated with an even stronger cross section
decrease for increasing energy, were seen for DR of HCl$^+$
\citep{Novotny:ApJ:2013} where a detailed discussion of this effect was also
given. 

We have excluded the possibility that these three  low-energy structures
are an experimental artifact due to
ion beam dragging. The concern is that if the electron beam is detuned
slightly from the cooling energy, such as for measurements at low collision
energies, then Coulomb forces between the beams can drag the ion beam to
match the velocity of the electrons. 
As a consequence the real collision energy would differ from that calculated by
Equation~\ref{eq:ed}. As discussed in \cite{Novotny:ApJ:2013}, this distorts the
curve of $\alpha_{\rm mb}$ at low collision energies and may appear as a low
energy peak. However, our experimental configuration with two electron beams
greatly minimizes this effect by enabling us to simultaneously cool the stored
ions with the Cooler while measuring DR data using the Target. Additionally, we
have performed a set of tests similar to \cite{Novotny:ApJ:2013} which further
excluded ion beam dragging as the origin for these features.

In the collision energy range between $\sim 0.05$~eV and $\sim 3$~eV a
significant enhancement is observed of the DR merged beams rate coefficient
compared to the $E^{-1}$ dependence for the direct DR, as can be seen in
Figure~\ref{fig:res_crosssec}. The rather narrow sub-structures in this energy
range (IV to XII, best seen in Fig.~\ref{fig:rateexpB}) suggest that indirect DR
is contributing in this energy range. 
Possible origins for the structure seen include the DR pathways via NH states
with electronically excited ion cores, the DR pathways via NH states with 
vibrationally excited ion cores, and the opening of new final
channels as we now describe.

At these energies similarly strong enhancements were seen for OH$^+$
\citep{AmitayOHp:PRA:1996}. There the comparison
with calculated data indicated that the enhancements stemmed from indirect DR
through neutral Rydberg resonances associated with bound, electronically excited
ion core molecular states. Comparable DR enhancements, possibly of the same
origin, have also been observed for CD$^+$ \citep{Forck:PRL:1994}, N$_2^+$
\citep{Peterson:JCP:1998}, and several other systems
\citep{Amitay:PRA:1996,Tanabe:JPB:1998,Padellec:JCP:1999, Novotny:JPCS:2009}.
For NH$^+$ neutral resonances of this type are expected to occur below the three
bound excited states $\rm A\,^2\Sigma^-$, $\rm B\,^2\Delta$, and $\rm
C\,^2\Sigma^+$, i.e., in the energy range of where the observed structure indeed
occurs in the measured DR rate coefficient. However, we are unaware of any
published potential energy curves for doubly excited bound neutral NH. Lacking
such calculations the observed structures cannot be uniquely assigned to
specific DR pathways via the neutral Rydberg series converging to the ionic A,
B, or C states. Nevertheless, as explained in Section~\ref{sec:DRpaths}, a
noticeable change in the DR rate is expected when scanning over a Rydberg series
limit converging to an ionic state. In the \NHp DR spectrum such a threshold
feature can be identified for the $v=0$ level of the A and B ionic states at
$\sim 2.7$~eV as a drop in the DR rate coefficient (structure XII).
Interestingly, no clear feature in $\alpha_{\rm mb}$ can be found at the
excitation energy of the $\rm C\,^2\Sigma^+$ state of \NHp.

In addition to the above DR via neutral Rydberg states converging to excited
electronic ionic states, some of the sub-structures in $\alpha_{\rm mb}$ above
$\sim 0.05$~eV may be also due to electron capture into neutral Rydberg series
converging to the vibrationally excited ${\rm X}\,^2\Pi (v\geq1)$ and ${\rm a}\,
^4\Sigma^- (v\geq1)$ \NHp states. The most clear feature suggesting an influence
of these excitation channels on the measured DR rate  coefficient is the sharp
drop in $\alpha_{\rm mb}$ at $\Ed\approx0.35$~eV (labeled by VIII). This edge
matches well to the $v=1$ levels of the $\rm X\,^2\Pi$ and $\rm a\, ^4\Sigma^-$
\NHp states. At higher energies, the upper edges of peaks X and XI may be
related to $v=2$ and $v=3$ levels of the $\rm X\,^2\Pi_{3/2}$ and $\rm a\,
^4\Sigma^-$ states of \NHp. However, between VIII and X lies peak IX, the upper
edge of which cannot be assigned to any \NHp vibrational level. This feature may
be due to DR resonances via doubly excited bound neutral states with an
electronically excited \NHp cores, as is discussed in the previous paragraph.
Feature VII possibly also has a similar origin.

Some of the structures seen in the $\sim 0.05$ to $\sim 3$~eV
energy range also seem to be related to the opening of new
final channels associated with higher excitation of the products. For example,
peak~IV at $\Ed\approx0.12$~eV matches very well with the threshold for product
excitation channel~4 of N and H from the \NHp ground state (see
Table~\ref{tab:channels}). According to \cite{Goldfield:JCP:1987} and
\cite{Owono:JCP:2007},  at least 6 NH states can predissociate to product
excitation channel~4. Taking all possible switching between these states into
account, a large number of predissociation pathways to channel 4 are available.
This is to be contrasted with the more endothermic product excitation channels
(5, 6, etc.) which do not show as good of a match with peaks in our experimental
DR spectrum. Moreover, the existence of additional peaks between the predicted
energy thresholds, such as feature IX, again suggests the existence of  more
complex DR pathways as discussed above.

A more general phenomenon of molecular DR, known for
most species, is the increase of the DR rate above $\sim 4$~eV
 with two broad peaks XIII and XIV at $\Ed\approx 7$~eV and
$10$~eV, respectively (see Fig.~\ref{fig:rateexpA}).
These reflect direct electron capture to NH
doubly excited unbound states with  $2\,^2\Pi$ and $2\,^4\Sigma^-$ \NHp ionic
cores. In Figures~\ref{fig:rateexpA} and \ref{fig:res_crosssec} we label the
respective vertical excitation energies for these two \NHp states by D and E.
We have investigated this phenomenon in more detail independently by the
fragment imaging technique \citep{Yang:JPB:2014}.

\subsection{Plasma rate coefficient}
Our experimentally derived DR plasma rate coefficient for \NHp differs
significantly from the currently recommended DR data. The only previous
experimental investigation for DR of \NHp is by
\cite{Mcgowan:jPRL:1979}. The results published in their original paper were
later corrected for an erroneous form factor
\citep{Mitchell:PR:1990,Mitchell:private:2012} yielding a plasma rate
coefficient of $\alpha_{\rm pl}^{\rm MG} = 4.3\e{-8}(300/T_{\rm
pl})^{0.5}$~cm$^3$s$^{-1}$ with an uncertainty of $\sim 15$\%.  Based on the
range of collision energies covered in the
experiment of \citeauthor{Mcgowan:jPRL:1979}, $\alpha_{\rm pl}^{\rm MG}$ is
valid for collisional plasma temperatures between approximately $100$~K and
$1000$~K. However, astrochemical models commonly extrapolate $\alpha_{\rm
pl}^{\rm MG}$ outside this temperature range by simply following the functional
expression given above. Comparing to our new results for DR of \NHp we find that
the magnitude of our rate coefficient is significantly larger than $\alpha_{\rm
pl}^{\rm MG}$ for all values of $T_{\rm pl}$. Additionally, we see that for
$T_{\rm pl}\lesssim1000$~K our $\alpha_{\rm pl}$ decreases with increasing
$T_{\rm pl}$ about $1.5$ times faster than $\alpha_{\rm pl}^{\rm MG}$. The 
unusual bump  in our thermal rate coefficient starting at $\sim 1000$~K and
reaching a maximum at $\sim 4000$~K
originates from the cross section enhancement at $E\approx0.05-3$~eV. 
As discussed above, this can be attributed to electron capture to doubly excited
states attached to electronically excited bound states of the \NHp ion.
Similar cross section enhancements were observed for other systems, such as
CH$^+$ and OH$^+$ \citep{Amitay:PRA:1996,AmitayOHp:PRA:1996}. The
corresponding high temperature increase in $\alpha_{\rm
pl}$ should thus be considered as a general DR feature which can not be
correctly described by the hyperbolic fit functions used by the astrochemical
databases. Instead a more general formula, such as equation
\ref{eq:plasmafitnew}, is needed to describe $\alpha_{\rm pl}$ for all $T_{\rm
pl}$ relevant for molecular ions.

Taking the ratio of $\alpha_{\rm
pl}/\alpha_{\rm pl}^{\rm MG}$ yields factors of 9.0, 5.8, 4.0, 2.8, and 2.3 at
$T=10$, 30, 100, 300, and 1000~K, respectively. These differences greatly
exceed the combined experimental error bars of \citeauthor{Mcgowan:jPRL:1979}
and our present results.
We attribute the differences between $\alpha_{\rm pl}$ and $\alpha_{\rm pl}^{\rm
MG}$ to the substantially different internal excitation of the \NHp ions in the
two experiments.
As discussed in Section~\ref{sec:NHpexc}, \NHp is expected to relax while
stored in TSR and reach thermal equilibrium with the 300~K temperatures of the
TSR chamber. Only rotational and fine-structure levels in the lowest two
electronic states are expected to  remain populated. On the other hand, in the
single-pass merged beam experiment of \citeauthor{Mcgowan:jPRL:1979} the flight
time from the ion source to the interaction region was $\sim 1~\mu$s. Thus, the
\NHp ions are expected to have been highly excited vibrationally, rotationally,
and electronically. Other experimental studies show that
internal excitation of the ions may affect significantly the DR rate coefficient
(e.g., \citealt{Amitay:Science:1998}, \citealt{Mccall:PRA:2004}).

Where no reliable DR data exist for diatomic molecules, astrochemical models
commonly  assume a ``typical'' rate coefficient of $\alpha_{\rm pl}^{\rm
di}\approx2.0\e{-7}\times(300/T)^{0.5}$~cm$^3\,$s$^{-1}$
\citep{Florescu:PR:2006}. As has been shown for CF$^+$
\citep{Novotny:JPB:2005} and HCl$^+$ \citep{Novotny:ApJ:2013},
$\alpha_{\rm pl}^{\rm di}$ does a poor job of matching experimentally derived
rate
coefficient $\alpha_{\rm pl}$ both in magnitude and temperature dependence. 
Taking the ratio of our new DR results to that commonly assumed we find
$\alpha_{\rm pl}/\alpha_{\rm pl}^{\rm di}=1.9$, 1.2, 0.9, 0.6, and 0.5 at
$T=10$, 30, 100, 300, and 1000~K, respectively.

\subsection{Astrochemical implications}
Although many attempts have been made to detect \NHp in the ISM, none have been
successful \citep{Snow:ASS:1979,Polehampton:MNRAS:2007,Hily-blant:AA:2010,
Persson:AA:2010,Persson:AA:2012,Benz:PCA:2013}. The non-detection of \NHp at the
sensitivity limits of the various observational techniques is in agreement with
existing astrochemical models (e.g., \citealt{Persson:AA:2010}). Our new data
predict that the \NHp destruction by DR at $T_{\rm pl}\lesssim50$~K is
significantly faster than previously assumed. For cold ISM environments at
$T_{\rm pl}=10$~K this enhancement reaches a factor of between 1.9 and 9.0
depending on the \NHp DR data used in the model (see
Figure~\ref{fig:rateplasmac2}). 

To estimate the relative effect of DR on the NH$^+$ abundance we compare the two
most important \NHp destruction channels: DR of \NHp (reaction \ref{eq:NHpDR})
and \NHp reacting with H$_2$ (reaction \ref{eq:NHpform}). Using the \NHp + H$_2$
reaction rates from \cite{Zymak:ApJ:2013} and our new \NHp DR data  we calculate
that DR is a competitive destruction channel at electron-to-H$_2$ density ratios
of $n_{\rm e}/n_{\rm H_2} \gtrsim 1.5\e{-5}$ and $\gtrsim 1.3\e{-3}$ at $T_{\rm
pl} = 10-100$~K, respectively. Such electron densities are predicted for diffuse
and translucent interstellar clouds \citep{Snow:ARAA:2006}. Our new DR data will
thus result in \NHp abundances significantly lower than currently predicted. As
a consequence, \NHp detection in the ISM is unlikely unless the current
detection limits can be lowered significantly. A reduced \NHp abundance will
also result in lower abundances for other nitrogen hydrides produced from \NHp
by reactions (\ref{eq:ammonia2}) and (\ref{eq:ammonia}). More precise
quantitative predictions will require that our new \NHp DR data be implemented
into current gas-phase and gas-grain nitrogen chemical models of the
interstellar clouds. That is, however, beyond the scope of our work.

Lastly, we note that even in our present storage ring experiment the NH$^+$ ions
are still internally excited to
$T_{\rm exc}\sim300$~K. In the cold ISM, however, the internal excitation is
expected to be significantly lower. This is because in the very low density ISM
environments the collision rate is much lower than the typical radiative decay
time \citep{Spitzer:ISM:1978}. Thus $T_{\rm exc}$ is generally even lower than
the kinetic temperature $T_{\rm pl}$ and most of the molecules are expected to
be in their rotational ground state. The response of the DR rate coefficients to
$T_{\rm exc}$ have been investigated only for light ions (e.g.,
\citealt{Amitay:PRA:1996}, \citealt{Zhaunerchyk:PRL:2007},
\citealt{Petrignani:PRA:2011}, \citealt{Schwalm:JPCS:2011}). From these few
studies we are unable to draw predictions for \NHp DR data with $T_{\rm
exc}<300$~K. Future studies at MPIK using the currently under construction
Cryogenic Storage Ring, which will have an internal ambient temperature of $\sim
10$~K, will be able to address this issue
\citep{Fadil:ACP:2006,Wolf:ACP:2006,Krantz:JPCS:2011,Von_hahn:NIMB:2011}.

\section{Summary}\label{sec:sum}
We have measured the absolute DR rate coefficient for \NHp in a merged beams
configuration at electron-ion collision energies up to 12~eV. For
astrophysical applications we have converted the experimental merged beams rate
coefficient to a cross section and a plasma rate coefficient. The resulting
plasma rate coefficient is faster in comparison to the DR data currently used
in most astrochemical models. Using the updated \NHp data we expect the \NHp
abundances will become even lower than currently predicted. This is in agreement
with the non-detection of \NHp in interstellar clouds. Our new \NHp data need to
be implemented in the state-of-the-art astrochemical models for quantitative
abundance predictions for nitrogen hydrides in general.

\acknowledgments
We thank the MPIK accelerator and TSR crews for their excellent support. We also
thank P.\ Pernot for stimulating discussions concerning the error analysis. ON
and DWS were supported in part by the NSF Division of Astronomical Sciences
Astronomy and Astrophysics Grants program and by the NASA Astronomy and Physics
Research and Analysis Program. DS acknowledges the support of the Weizmann
Institute of Science through the Joseph Meyerhoff program. The work is supported
in part by the German-Israeli Foundation for Scientific Research [GIF under
contract nr. I-900-231.7/2005]. WG acknowledges partial support by the COST
Action CM0805: ``The Chemical Cosmos: Understanding Chemistry in Astronomical
Environments''. BY thanks for the support from MPG, the National Basic Research
Program of China (Grant No. 2010CB832901), and the Joint Funds of the National
Natural Science Foundation of China (Grant No. U1332206). The work is supported
in part by the DFG Priority Program 1573 ``Physics of the Interstellar Medium". 


\bibliography{NH+_DR_ApJ}


\begin{deluxetable}{crrc}
\tablecolumns{4} 
\tablewidth{0pc} 
\tablecaption{ Product excitation channels for DR of ground state \NHp
at an electron-ion collision energy of $E=0$~eV. 
		\label{tab:channels}
		}
\tablehead{ \colhead{ID} & \colhead{H state} & \colhead{N state}  &
\colhead{KER [eV]}
}
\startdata
1 & $1s\ \rm ^2S_{1/2}$	& $2s^2 2p^3\ \rm ^4S^o_{3/2}$	& $10.074$
\\
2 & $1s\ \rm ^2S_{1/2}$	& $2s^2 2p^3\ \rm ^2D^o_{5/2}$	& $7.690$ \\
3 & $1s\ \rm ^2S_{1/2}$	& $2s^2 2p^3\ \rm ^2P^o_{1/2}$	& $6.498$ \\
4 & $2p\ \rm ^2P^o_{1/2}$ &$2s^2 2p^3\ \rm ^4S^o_{3/2}$	&
$-0.125$ \\
  & $2s\ \rm ^2S_{1/2}$ & $2s^2 2p^3\ \rm ^4S^o_{3/2}$	&
$-0.125$ \\
5 & $1s\ \rm ^2S_{1/2}$	& $2s^2 2p^2(^3{\rm P})3s\ {\rm ^4P_{1/2}}$	&
$-0.252$ \\
6 & $1s\ \rm ^2S_{1/2}$	& $2s^2 2p^2(^3{\rm P})3s\ \rm ^2P_{1/2}$	&
$-0.606$ \\
7 & $1s\ \rm ^2S_{1/2}$	& $2s^2 2p^4\ {\rm ^4P_{5/2}}$	& $-0.850$ \\
8 & $1s\ \rm ^2S_{1/2}$	& $2s^2 2p^2(^3{\rm P})3p\ {\rm ^2S^o_{1/2}}$	&
$-1.529$ \\
\enddata
\tablecomments{\doublespace
For each electronic state of H or N only the lowest fine-structure level is
listed. The fine structure splitting for the omitted levels is less than 10~meV
each. Negative KER values indicate channels energetically closed at $E=0$~eV,
but which become energetically allowed for $E\geq-{\rm KER}$. Channel 4 has two
possible hydrogen configurations which differ energetically by less than
$5~\mu$eV.}
\end{deluxetable}


\begin{deluxetable}{ccc}
\tablecolumns{3} 
\tablewidth{0pc} 
\tablecaption{Experimental merged beams rate coefficient $\alpha_{\rm mb}$
for DR of NH$^+$. \label{tab:data:MB}
}
\tablehead{ \colhead{$E_{\rm d}$} & \colhead{$\alpha_{\rm mb}$}  &
\colhead{Statistical error}\\
\colhead{(eV)} & \colhead{(cm$^3\,$s$^{-1}$)}  &
\colhead{(cm$^3\,$s$^{-1}$)}
}
\startdata
1.71(-5) & 1.79(-6) & 6.63(-8) \\
2.43(-5) & 1.82(-6) & 7.88(-8) \\
3.43(-5) & 1.67(-6) & 5.50(-8) \\
4.85(-5) & 1.55(-6) & 5.68(-8) \\
6.83(-5) & 1.47(-6) & 6.60(-8) \\
9.60(-5) & 1.25(-6) & 3.98(-8) \\
1.34(-4) & 1.06(-6) & 3.11(-8) \\
1.87(-4) & 8.38(-7) & 3.39(-8) \\
2.57(-4) & 7.69(-7) & 2.96(-8) \\
3.50(-4) & 6.36(-7) & 1.98(-8) \\
4.70(-4) & 5.32(-7) & 1.52(-8) \\
6.18(-4) & 4.83(-7) & 1.39(-8) \\
7.92(-4) & 4.52(-7) & 1.30(-8) \\
9.85(-4) & 4.41(-7) & 1.46(-8) \\
1.19(-3) & 3.79(-7) & 1.16(-8) \\
1.43(-3) & 3.95(-7) & 1.91(-8) \\
1.71(-3) & 3.69(-7) & 9.75(-9) \\
2.04(-3) & 3.40(-7) & 9.53(-9) \\
2.43(-3) & 2.98(-7) & 8.33(-9) \\
2.89(-3) & 2.91(-7) & 7.57(-9) \\
3.42(-3) & 2.69(-7) & 4.46(-9) \\
4.04(-3) & 2.51(-7) & 2.38(-9) \\
4.75(-3) & 2.36(-7) & 2.92(-9) \\
5.55(-3) & 2.13(-7) & 1.97(-9) \\
6.45(-3) & 1.90(-7) & 1.89(-9) \\
7.45(-3) & 1.71(-7) & 1.81(-9) \\
8.55(-3) & 1.55(-7) & 1.80(-9) \\
9.73(-3) & 1.42(-7) & 1.70(-9) \\
1.10(-2) & 1.30(-7) & 1.77(-9) \\
1.23(-2) & 1.24(-7) & 1.76(-9) \\
1.36(-2) & 1.20(-7) & 1.79(-9) \\
1.49(-2) & 1.10(-7) & 1.66(-9) \\
1.62(-2) & 1.06(-7) & 1.93(-9) \\
1.74(-2) & 9.92(-8) & 1.96(-9) \\
1.86(-2) & 9.17(-8) & 1.92(-9) \\
1.96(-2) & 8.82(-8) & 1.93(-9) \\
2.06(-2) & 8.24(-8) & 1.84(-9) \\
2.15(-2) & 8.40(-8) & 2.25(-9) \\
2.25(-2) & 7.93(-8) & 1.59(-9) \\
2.35(-2) & 7.57(-8) & 1.55(-9) \\
2.45(-2) & 7.27(-8) & 2.10(-9) \\
2.56(-2) & 7.01(-8) & 1.54(-9) \\
2.67(-2) & 6.53(-8) & 1.49(-9) \\
2.79(-2) & 6.27(-8) & 1.53(-9) \\
2.91(-2) & 6.71(-8) & 2.68(-9) \\
3.04(-2) & 5.91(-8) & 1.87(-9) \\
3.17(-2) & 5.42(-8) & 2.02(-9) \\
3.31(-2) & 5.46(-8) & 1.85(-9) \\
3.45(-2) & 4.88(-8) & 1.94(-9) \\
3.60(-2) & 4.66(-8) & 1.77(-9) \\
3.76(-2) & 4.28(-8) & 2.07(-9) \\
3.92(-2) & 4.36(-8) & 1.82(-9) \\
4.08(-2) & 4.08(-8) & 7.19(-10) \\
4.25(-2) & 3.75(-8) & 1.46(-9) \\
4.43(-2) & 3.95(-8) & 6.27(-10) \\
4.62(-2) & 3.38(-8) & 6.79(-10) \\
4.81(-2) & 3.30(-8) & 6.40(-10) \\
5.01(-2) & 2.97(-8) & 5.24(-10) \\
5.21(-2) & 2.75(-8) & 1.07(-9) \\
5.42(-2) & 2.63(-8) & 4.47(-10) \\
5.64(-2) & 2.51(-8) & 1.08(-9) \\
5.86(-2) & 2.49(-8) & 5.02(-10) \\
6.10(-2) & 2.43(-8) & 5.84(-10) \\
6.34(-2) & 2.49(-8) & 6.04(-10) \\
6.58(-2) & 2.41(-8) & 4.46(-10) \\
6.84(-2) & 2.48(-8) & 1.44(-9) \\
7.10(-2) & 2.61(-8) & 6.10(-10) \\
7.37(-2) & 2.70(-8) & 5.51(-10) \\
7.93(-2) & 2.83(-8) & 6.14(-10) \\
8.22(-2) & 2.95(-8) & 9.95(-10) \\
8.52(-2) & 2.92(-8) & 7.95(-10) \\
8.83(-2) & 3.01(-8) & 5.90(-10) \\
9.46(-2) & 3.11(-8) & 6.15(-10) \\
9.79(-2) & 3.19(-8) & 6.21(-10) \\
1.05(-1) & 3.34(-8) & 6.37(-10) \\
1.08(-1) & 3.32(-8) & 6.02(-10) \\
1.15(-1) & 3.85(-8) & 6.33(-10) \\
1.19(-1) & 4.11(-8) & 6.61(-10) \\
1.26(-1) & 4.18(-8) & 6.64(-10) \\
1.30(-1) & 4.04(-8) & 7.20(-10) \\
1.38(-1) & 3.88(-8) & 7.17(-10) \\
1.42(-1) & 3.78(-8) & 6.28(-10) \\
1.50(-1) & 3.85(-8) & 6.71(-10) \\
1.54(-1) & 3.66(-8) & 6.54(-10) \\
1.62(-1) & 3.38(-8) & 6.34(-10) \\
1.66(-1) & 3.42(-8) & 6.66(-10) \\
1.74(-1) & 3.14(-8) & 6.01(-10) \\
1.79(-1) & 3.39(-8) & 7.47(-10) \\
1.83(-1) & 3.31(-8) & 1.04(-9) \\
1.87(-1) & 3.58(-8) & 6.10(-10) \\
1.95(-1) & 3.83(-8) & 5.90(-10) \\
1.99(-1) & 3.73(-8) & 8.44(-10) \\
2.03(-1) & 3.81(-8) & 9.11(-10) \\
2.07(-1) & 3.76(-8) & 5.55(-10) \\
2.15(-1) & 3.54(-8) & 5.61(-10) \\
2.24(-1) & 3.84(-8) & 5.78(-10) \\
2.29(-1) & 4.08(-8) & 6.07(-10) \\
2.39(-1) & 4.63(-8) & 6.19(-10) \\
2.44(-1) & 4.56(-8) & 8.53(-10) \\
2.49(-1) & 4.94(-8) & 1.03(-9) \\
2.54(-1) & 4.49(-8) & 6.60(-10) \\
2.59(-1) & 4.29(-8) & 6.96(-10) \\
2.65(-1) & 4.37(-8) & 1.31(-9) \\
2.71(-1) & 4.14(-8) & 5.81(-10) \\
2.76(-1) & 4.39(-8) & 7.29(-10) \\
2.89(-1) & 4.50(-8) & 6.31(-10) \\
2.95(-1) & 5.08(-8) & 6.64(-10) \\
3.01(-1) & 5.36(-8) & 7.23(-10) \\
3.15(-1) & 5.20(-8) & 7.63(-10) \\
3.22(-1) & 4.75(-8) & 6.70(-10) \\
3.30(-1) & 4.45(-8) & 6.44(-10) \\
3.37(-1) & 3.85(-8) & 6.34(-10) \\
3.45(-1) & 3.55(-8) & 5.82(-10) \\
3.61(-1) & 3.27(-8) & 6.30(-10) \\
3.70(-1) & 3.03(-8) & 5.45(-10) \\
3.79(-1) & 2.47(-8) & 4.83(-10) \\
3.88(-1) & 2.13(-8) & 5.40(-10) \\
3.97(-1) & 1.89(-8) & 4.82(-10) \\
4.07(-1) & 1.72(-8) & 4.92(-10) \\
4.17(-1) & 1.52(-8) & 4.88(-10) \\
4.28(-1) & 1.58(-8) & 5.15(-10) \\
4.39(-1) & 1.71(-8) & 4.78(-10) \\
4.50(-1) & 1.98(-8) & 4.89(-10) \\
4.61(-1) & 2.21(-8) & 5.24(-10) \\
4.73(-1) & 2.33(-8) & 3.95(-10) \\
4.86(-1) & 2.55(-8) & 4.97(-10) \\
4.99(-1) & 2.64(-8) & 5.55(-10) \\
5.12(-1) & 3.15(-8) & 5.44(-10) \\
5.26(-1) & 3.27(-8) & 5.93(-10) \\
5.40(-1) & 2.81(-8) & 4.07(-10) \\
5.56(-1) & 2.42(-8) & 3.89(-10) \\
5.71(-1) & 2.16(-8) & 5.42(-10) \\
5.88(-1) & 1.62(-8) & 2.79(-10) \\
6.05(-1) & 1.44(-8) & 3.64(-10) \\
6.22(-1) & 1.65(-8) & 3.05(-10) \\
6.41(-1) & 1.96(-8) & 3.17(-10) \\
6.60(-1) & 2.33(-8) & 4.19(-10) \\
6.80(-1) & 2.36(-8) & 3.40(-10) \\
7.01(-1) & 2.07(-8) & 3.13(-10) \\
7.24(-1) & 1.62(-8) & 3.71(-10) \\
7.47(-1) & 1.74(-8) & 2.89(-10) \\
7.71(-1) & 1.86(-8) & 2.84(-10) \\
7.97(-1) & 1.81(-8) & 3.26(-10) \\
8.24(-1) & 1.98(-8) & 3.71(-10) \\
8.52(-1) & 2.04(-8) & 2.61(-10) \\
8.82(-1) & 2.01(-8) & 2.47(-10) \\
9.13(-1) & 1.75(-8) & 2.53(-10) \\
9.47(-1) & 1.45(-8) & 2.44(-10) \\
9.82(-1) & 1.24(-8) & 3.22(-10) \\
1.02(+0) & 1.23(-8) & 5.83(-10) \\
1.06(+0) & 9.97(-9) & 5.37(-10) \\
1.10(+0) & 9.40(-9) & 5.13(-10) \\
1.15(+0) & 9.01(-9) & 5.26(-10) \\
1.19(+0) & 6.61(-9) & 5.12(-10) \\
1.25(+0) & 6.38(-9) & 4.75(-10) \\
1.30(+0) & 6.60(-9) & 3.42(-10) \\
1.36(+0) & 4.81(-9) & 4.26(-10) \\
1.42(+0) & 5.58(-9) & 3.34(-10) \\
1.49(+0) & 3.87(-9) & 4.46(-10) \\
1.57(+0) & 4.46(-9) & 3.10(-10) \\
1.64(+0) & 3.86(-9) & 4.27(-10) \\
1.73(+0) & 3.37(-9) & 2.77(-10) \\
1.81(+0) & 3.08(-9) & 2.81(-10) \\
1.90(+0) & 2.94(-9) & 2.76(-10) \\
1.99(+0) & 3.90(-9) & 4.20(-10) \\
2.09(+0) & 2.80(-9) & 2.73(-10) \\
2.19(+0) & 2.91(-9) & 2.95(-10) \\
2.30(+0) & 2.86(-9) & 2.86(-10) \\
2.41(+0) & 2.59(-9) & 2.80(-10) \\
2.53(+0) & 3.46(-9) & 2.98(-10) \\
2.65(+0) & 3.27(-9) & 3.16(-10) \\
2.78(+0) & 2.78(-9) & 3.10(-10) \\
2.91(+0) & 1.61(-9) & 2.12(-10) \\
3.05(+0) & 1.10(-9) & 2.79(-10) \\
3.19(+0) & 5.85(-10) & 2.60(-10) \\
3.34(+0) & 5.23(-10) & 2.81(-10) \\
3.50(+0) & 5.74(-10) & 2.16(-10) \\
3.66(+0) & 1.06(-9) & 2.95(-10) \\
3.83(+0) & 5.28(-10) & 2.17(-10) \\
4.00(+0) & 7.56(-10) & 2.77(-10) \\
4.19(+0) & 1.12(-9) & 2.30(-10) \\
4.38(+0) & 1.16(-9) & 2.86(-10) \\
4.57(+0) & 1.33(-9) & 2.36(-10) \\
4.77(+0) & 2.09(-9) & 3.03(-10) \\
4.98(+0) & 1.65(-9) & 2.37(-10) \\
5.20(+0) & 1.94(-9) & 2.53(-10) \\
5.43(+0) & 2.38(-9) & 1.81(-10) \\
5.66(+0) & 2.37(-9) & 2.35(-10) \\
5.90(+0) & 2.94(-9) & 1.97(-10) \\
6.15(+0) & 2.91(-9) & 1.96(-10) \\
6.40(+0) & 3.85(-9) & 2.25(-10) \\
6.67(+0) & 3.91(-9) & 3.22(-10) \\
6.94(+0) & 4.16(-9) & 3.56(-10) \\
7.22(+0) & 4.50(-9) & 3.65(-10) \\
7.51(+0) & 4.09(-9) & 3.68(-10) \\
7.81(+0) & 4.31(-9) & 3.10(-10) \\
8.13(+0) & 5.63(-9) & 3.74(-10) \\
8.46(+0) & 5.32(-9) & 4.01(-10) \\
8.81(+0) & 6.61(-9) & 3.79(-10) \\
9.18(+0) & 7.66(-9) & 3.81(-10) \\
9.56(+0) & 6.94(-9) & 4.24(-10) \\
9.97(+0) & 8.10(-9) & 4.21(-10) \\
1.04(+1) & 7.00(-9) & 3.69(-10) \\
1.08(+1) & 7.15(-9) & 3.24(-10) \\
1.13(+1) & 6.73(-9) & 3.88(-10) \\
1.18(+1) & 6.20(-9) & 3.48(-10) \\
\enddata
\tablecomments{\doublespace
The format $x(y)$ signifies $x\e{y}$. There is an additional systematic
uncertainty of 8.4\% in the merged beams rate coefficient (see text).
}
\end{deluxetable}

\clearpage

\begin{deluxetable}{ccccccc}
\tablecolumns{3} 
\tablewidth{0pc} 
\tablecaption{Cross section $\sigma$
for DR of NH$^+$. \label{tab:data:CS}
}
\tablehead{ \colhead{$E_{\rm center}$} & \colhead{$E_{\rm width}$} &
\colhead{$\sigma$}  
& \colhead {$U_{\rm stat}^{\rm lo}$} & \colhead{$U_{\rm stat}^{\rm hi}$}
& \colhead {$U_{\rm syst}^{\rm lo}$} & \colhead{$U_{\rm syst}^{\rm hi}$}\\
\colhead{(eV)} & \colhead{(eV)} & \colhead{(cm$^2$)} 
& \colhead{(cm$^2$)} & \colhead{(cm$^2$)} & \colhead{(cm$^2$)} &
\colhead{(cm$^2$)}
}
\startdata
1.65(-5) & 3.30(-5) & 1.28(-10) & -4.38(-11) & +4.16(-11) & -2.76(-11) &
+2.15(-10) \\
8.36(-5) & 1.01(-4) & 3.86(-11) & -5.16(-12) & +5.58(-12) & -7.80(-12) &
+8.19(-12) \\
2.52(-4) & 2.35(-4) & 4.43(-12) & -6.96(-13) & +6.58(-13) & -1.77(-12) &
+6.87(-13) \\
6.13(-4) & 4.88(-4) & 7.86(-13) & -9.02(-14) & +9.17(-14) & -1.37(-13) &
+1.13(-13) \\
1.35(-3) & 9.87(-4) & 3.12(-13) & -1.89(-14) & +1.77(-14) & -3.09(-14) &
+3.14(-14) \\
2.52(-3) & 1.35(-3) & 1.67(-13) & -6.70(-15) & +6.79(-15) & -1.64(-14) &
+1.63(-14) \\
3.93(-3) & 1.48(-3) & 9.62(-14) & -3.31(-15) & +2.88(-15) & -1.02(-14) &
+8.32(-15) \\
5.48(-3) & 1.61(-3) & 7.45(-14) & -1.88(-15) & +1.90(-15) & -6.90(-15) &
+8.27(-15) \\
7.16(-3) & 1.75(-3) & 4.95(-14) & -1.26(-15) & +1.31(-15) & -5.36(-15) &
+4.45(-15) \\
8.97(-3) & 1.88(-3) & 3.69(-14) & -1.21(-15) & +1.03(-15) & -3.26(-15) &
+3.77(-15) \\
1.09(-2) & 2.01(-3) & 2.72(-14) & -9.67(-16) & +9.75(-16) & -2.92(-15) &
+2.33(-15) \\
1.30(-2) & 2.15(-3) & 2.26(-14) & -7.54(-16) & +8.89(-16) & -1.93(-15) &
+2.08(-15) \\
1.52(-2) & 2.28(-3) & 1.92(-14) & -7.69(-16) & +6.85(-16) & -1.70(-15) &
+1.66(-15) \\
1.76(-2) & 2.42(-3) & 1.65(-14) & -6.76(-16) & +6.70(-16) & -1.44(-15) &
+1.55(-15) \\
2.01(-2) & 2.55(-3) & 1.25(-14) & -5.19(-16) & +5.21(-16) & -1.19(-15) &
+1.09(-15) \\
2.27(-2) & 2.69(-3) & 1.10(-14) & -4.55(-16) & +5.13(-16) & -9.54(-16) &
+9.63(-16) \\
2.54(-2) & 2.82(-3) & 9.35(-15) & -4.00(-16) & +4.28(-16) & -8.38(-16) &
+9.29(-16) \\
2.83(-2) & 2.96(-3) & 7.35(-15) & -4.21(-16) & +3.27(-16) & -9.52(-16) &
+6.48(-16) \\
3.13(-2) & 3.09(-3) & 6.92(-15) & -4.04(-16) & +4.53(-16) & -6.32(-16) &
+9.05(-16) \\
3.45(-2) & 3.23(-3) & 5.45(-15) & -3.72(-16) & +4.13(-16) & -6.10(-16) &
+5.21(-16) \\
3.78(-2) & 3.36(-3) & 4.43(-15) & -3.42(-16) & +3.00(-16) & -4.35(-16) &
+4.55(-16) \\
4.12(-2) & 3.50(-3) & 3.65(-15) & -1.68(-16) & +1.62(-16) & -5.36(-16) &
+3.48(-16) \\
4.48(-2) & 3.64(-3) & 3.71(-15) & -1.57(-16) & +1.32(-16) & -3.65(-16) &
+6.00(-16) \\
4.85(-2) & 3.77(-3) & 2.70(-15) & -1.26(-16) & +1.35(-16) & -4.08(-16) &
+2.73(-16) \\
5.24(-2) & 3.91(-3) & 2.23(-15) & -1.28(-16) & +1.13(-16) & -2.36(-16) &
+2.74(-16) \\
5.63(-2) & 4.05(-3) & 1.82(-15) & -9.06(-17) & +9.43(-17) & -2.21(-16) &
+1.98(-16) \\
6.04(-2) & 4.18(-3) & 1.65(-15) & -8.59(-17) & +8.45(-17) & -1.84(-16) &
+1.82(-16) \\
6.47(-2) & 4.32(-3) & 1.64(-15) & -1.08(-16) & +8.91(-17) & -1.84(-16) &
+2.06(-16) \\
6.91(-2) & 4.45(-3) & 1.40(-15) & -1.06(-16) & +1.15(-16) & -2.46(-16) &
+1.56(-16) \\
7.36(-2) & 4.59(-3) & 1.71(-15) & -1.09(-16) & +1.10(-16) & -1.84(-16) &
+2.81(-16) \\
7.83(-2) & 4.73(-3) & 1.57(-15) & -1.68(-16) & +1.74(-16) & -2.40(-16) &
+1.72(-16) \\
8.31(-2) & 4.86(-3) & 1.78(-15) & -1.41(-16) & +1.30(-16) & -1.88(-16) &
+2.62(-16) \\
8.80(-2) & 5.00(-3) & 1.64(-15) & -1.07(-16) & +1.21(-16) & -2.80(-16) &
+1.72(-16) \\
9.31(-2) & 5.14(-3) & 1.80(-15) & -1.62(-16) & +1.38(-16) & -1.85(-16) &
+3.15(-16) \\
9.83(-2) & 5.27(-3) & 1.65(-15) & -8.93(-17) & +9.83(-17) & -3.37(-16) &
+1.70(-16) \\
1.04(-1) & 5.41(-3) & 1.91(-15) & -1.28(-16) & +1.22(-16) & -1.95(-16) &
+3.56(-16) \\
1.09(-1) & 5.55(-3) & 1.63(-15) & -7.99(-17) & +7.66(-17) & -3.02(-16) &
+1.65(-16) \\
1.15(-1) & 5.69(-3) & 1.90(-15) & -1.04(-16) & +9.87(-17) & -1.83(-16) &
+2.02(-16) \\
1.20(-1) & 5.82(-3) & 2.18(-15) & -7.33(-17) & +8.23(-17) & -2.06(-16) &
+2.06(-16) \\
1.26(-1) & 5.96(-3) & 2.21(-15) & -9.81(-17) & +8.17(-17) & -2.09(-16) &
+2.61(-16) \\
1.32(-1) & 6.10(-3) & 1.98(-15) & -7.74(-17) & +8.00(-17) & -2.56(-16) &
+1.91(-16) \\
1.39(-1) & 6.23(-3) & 1.92(-15) & -7.52(-17) & +8.76(-17) & -1.84(-16) &
+3.06(-16) \\
1.45(-1) & 6.37(-3) & 1.73(-15) & -8.21(-17) & +7.22(-17) & -3.74(-16) &
+1.67(-16) \\
1.51(-1) & 6.51(-3) & 1.88(-15) & -7.06(-17) & +7.73(-17) & -1.82(-16) &
+4.51(-16) \\
1.58(-1) & 6.64(-3) & 1.53(-15) & -8.57(-17) & +8.73(-17) & -4.03(-16) &
+1.52(-16) \\
1.65(-1) & 6.78(-3) & 1.52(-15) & -6.82(-17) & +6.45(-17) & -1.51(-16) &
+3.14(-16) \\
1.71(-1) & 6.92(-3) & 1.35(-15) & -1.07(-16) & +9.58(-17) & -2.62(-16) &
+1.36(-16) \\
1.78(-1) & 7.06(-3) & 1.32(-15) & -6.06(-17) & +6.17(-17) & -1.32(-16) &
+1.73(-16) \\
1.86(-1) & 7.19(-3) & 1.36(-15) & -6.86(-17) & +7.40(-17) & -1.87(-16) &
+1.34(-16) \\
1.93(-1) & 7.33(-3) & 1.62(-15) & -9.83(-17) & +8.85(-17) & -1.57(-16) &
+2.00(-16) \\
2.00(-1) & 7.47(-3) & 1.48(-15) & -6.17(-17) & +7.17(-17) & -1.74(-16) &
+1.43(-16) \\
2.08(-1) & 7.61(-3) & 1.53(-15) & -4.42(-17) & +4.28(-17) & -1.48(-16) &
+1.86(-16) \\
2.15(-1) & 7.74(-3) & 1.30(-15) & -4.92(-17) & +5.15(-17) & -1.46(-16) &
+1.27(-16) \\
2.23(-1) & 7.88(-3) & 1.37(-15) & -5.99(-17) & +6.45(-17) & -1.47(-16) &
+1.32(-16) \\
2.31(-1) & 8.02(-3) & 1.53(-15) & -5.01(-17) & +4.76(-17) & -1.44(-16) &
+1.55(-16) \\
2.39(-1) & 8.16(-3) & 1.72(-15) & -5.16(-17) & +4.74(-17) & -1.64(-16) &
+1.59(-16) \\
2.47(-1) & 8.29(-3) & 1.80(-15) & -5.25(-17) & +6.46(-17) & -1.65(-16) &
+1.86(-16) \\
2.56(-1) & 8.43(-3) & 1.64(-15) & -5.00(-17) & +4.78(-17) & -1.53(-16) &
+1.53(-16) \\
2.64(-1) & 8.57(-3) & 1.50(-15) & -5.37(-17) & +6.41(-17) & -1.43(-16) &
+1.41(-16) \\
2.73(-1) & 8.71(-3) & 1.40(-15) & -4.23(-17) & +4.74(-17) & -1.40(-16) &
+1.32(-16) \\
2.82(-1) & 8.84(-3) & 1.62(-15) & -7.37(-17) & +6.95(-17) & -1.51(-16) &
+1.73(-16) \\
2.91(-1) & 8.98(-3) & 1.46(-15) & -3.97(-17) & +4.09(-17) & -2.29(-16) &
+1.36(-16) \\
3.00(-1) & 9.12(-3) & 1.89(-15) & -4.32(-17) & +4.50(-17) & -1.71(-16) &
+2.16(-16) \\
3.11(-1) & 1.44(-2) & 1.86(-15) & -5.21(-17) & +5.21(-17) & -1.69(-16) &
+1.71(-16) \\
3.22(-1) & 7.25(-3) & 1.57(-15) & -5.91(-17) & +6.68(-17) & -1.64(-16) &
+1.47(-16) \\
3.30(-1) & 7.47(-3) & 1.58(-15) & -6.43(-17) & +5.95(-17) & -1.51(-16) &
+1.99(-16) \\
3.37(-1) & 7.70(-3) & 1.26(-15) & -5.21(-17) & +5.77(-17) & -1.59(-16) &
+1.23(-16) \\
3.47(-1) & 1.21(-2) & 1.12(-15) & -3.15(-17) & +2.61(-17) & -1.10(-16) &
+1.12(-16) \\
3.59(-1) & 1.25(-2) & 1.03(-15) & -3.69(-17) & +4.18(-17) & -1.25(-16) &
+1.02(-16) \\
3.70(-1) & 8.71(-3) & 1.01(-15) & -3.98(-17) & +3.71(-17) & -1.02(-16) &
+2.05(-16) \\
3.79(-1) & 9.00(-3) & 7.46(-16) & -3.44(-17) & +3.23(-17) & -1.67(-16) &
+8.03(-17) \\
3.88(-1) & 9.29(-3) & 6.29(-16) & -3.33(-17) & +3.55(-17) & -7.22(-17) &
+1.04(-16) \\
3.97(-1) & 9.60(-3) & 5.31(-16) & -2.94(-17) & +2.98(-17) & -9.56(-17) &
+6.39(-17) \\
4.07(-1) & 9.92(-3) & 4.92(-16) & -3.00(-17) & +3.05(-17) & -6.35(-17) &
+9.55(-17) \\
4.17(-1) & 1.03(-2) & 3.75(-16) & -2.69(-17) & +2.84(-17) & -8.89(-17) &
+5.08(-17) \\
4.28(-1) & 1.06(-2) & 3.98(-16) & -2.80(-17) & +2.77(-17) & -5.19(-17) &
+5.81(-17) \\
4.39(-1) & 1.10(-2) & 4.14(-16) & -2.61(-17) & +2.50(-17) & -5.63(-17) &
+5.32(-17) \\
4.50(-1) & 1.14(-2) & 4.96(-16) & -2.49(-17) & +2.29(-17) & -6.13(-17) &
+5.87(-17) \\
4.61(-1) & 1.18(-2) & 5.75(-16) & -2.43(-17) & +2.40(-17) & -6.45(-17) &
+8.27(-17) \\
4.73(-1) & 1.22(-2) & 5.80(-16) & -1.96(-17) & +1.96(-17) & -8.99(-17) &
+6.52(-17) \\
4.86(-1) & 1.27(-2) & 6.68(-16) & -2.21(-17) & +2.14(-17) & -7.03(-17) &
+9.85(-17) \\
4.99(-1) & 1.32(-2) & 6.39(-16) & -2.36(-17) & +2.31(-17) & -1.01(-16) &
+6.57(-17) \\
5.12(-1) & 1.37(-2) & 8.15(-16) & -2.14(-17) & +2.25(-17) & -8.12(-17) &
+8.48(-17) \\
5.26(-1) & 1.42(-2) & 8.81(-16) & -2.38(-17) & +2.32(-17) & -8.77(-17) &
+9.41(-17) \\
5.41(-1) & 1.48(-2) & 7.26(-16) & -1.56(-17) & +1.55(-17) & -7.58(-17) &
+7.62(-17) \\
5.56(-1) & 1.54(-2) & 6.03(-16) & -1.58(-17) & +1.38(-17) & -7.16(-17) &
+6.46(-17) \\
5.71(-1) & 1.60(-2) & 5.49(-16) & -1.68(-17) & +1.82(-17) & -6.24(-17) &
+7.36(-17) \\
5.88(-1) & 1.67(-2) & 3.74(-16) & -1.04(-17) & +9.33(-18) & -5.15(-17) &
+4.90(-17) \\
6.05(-1) & 1.74(-2) & 3.05(-16) & -1.13(-17) & +1.18(-17) & -4.77(-17) &
+4.43(-17) \\
6.22(-1) & 1.81(-2) & 3.55(-16) & -9.79(-18) & +8.36(-18) & -4.62(-17) &
+4.62(-17) \\
6.41(-1) & 1.89(-2) & 4.26(-16) & -9.19(-18) & +1.00(-17) & -5.10(-17) &
+4.96(-17) \\
6.60(-1) & 1.98(-2) & 5.28(-16) & -1.38(-17) & +1.18(-17) & -5.71(-17) &
+5.79(-17) \\
6.81(-1) & 2.07(-2) & 5.38(-16) & -9.75(-18) & +9.36(-18) & -5.88(-17) &
+5.92(-17) \\
7.02(-1) & 2.16(-2) & 4.65(-16) & -9.49(-18) & +8.22(-18) & -5.33(-17) &
+5.58(-17) \\
7.24(-1) & 2.27(-2) & 3.32(-16) & -1.03(-17) & +1.04(-17) & -4.79(-17) &
+4.35(-17) \\
7.47(-1) & 2.38(-2) & 3.59(-16) & -7.03(-18) & +7.90(-18) & -4.58(-17) &
+4.58(-17) \\
7.71(-1) & 2.50(-2) & 3.85(-16) & -7.09(-18) & +7.62(-18) & -4.69(-17) &
+4.72(-17) \\
7.97(-1) & 2.63(-2) & 3.63(-16) & -8.34(-18) & +8.15(-18) & -4.63(-17) &
+4.58(-17) \\
8.24(-1) & 2.76(-2) & 4.01(-16) & -7.89(-18) & +8.58(-18) & -4.63(-17) &
+4.63(-17) \\
8.52(-1) & 2.91(-2) & 4.10(-16) & -5.70(-18) & +6.40(-18) & -4.81(-17) &
+4.81(-17) \\
8.82(-1) & 3.07(-2) & 4.03(-16) & -5.69(-18) & +5.82(-18) & -4.79(-17) &
+4.82(-17) \\
9.14(-1) & 3.24(-2) & 3.45(-16) & -5.53(-18) & +5.69(-18) & -4.31(-17) &
+4.32(-17) \\
9.47(-1) & 3.43(-2) & 2.77(-16) & -5.18(-18) & +5.31(-18) & -3.95(-17) &
+3.95(-17) \\
9.83(-1) & 3.63(-2) & 2.29(-16) & -6.65(-18) & +6.83(-18) & -3.72(-17) &
+3.70(-17) \\
1.02(+0) & 3.85(-2) & 2.26(-16) & -1.15(-17) & +1.33(-17) & -2.16(-17) &
+2.23(-17) \\
1.06(+0) & 4.09(-2) & 1.77(-16) & -9.88(-18) & +1.01(-17) & -1.74(-17) &
+1.74(-17) \\
1.10(+0) & 4.36(-2) & 1.64(-16) & -1.06(-17) & +9.14(-18) & -1.62(-17) &
+1.62(-17) \\
1.15(+0) & 4.64(-2) & 1.56(-16) & -1.09(-17) & +1.03(-17) & -1.56(-17) &
+1.57(-17) \\
1.19(+0) & 4.96(-2) & 1.09(-16) & -1.02(-17) & +9.20(-18) & -1.15(-17) &
+1.15(-17) \\
1.25(+0) & 5.31(-2) & 1.03(-16) & -9.35(-18) & +8.48(-18) & -1.09(-17) &
+1.09(-17) \\
1.30(+0) & 5.70(-2) & 1.06(-16) & -6.24(-18) & +6.42(-18) & -1.13(-17) &
+1.15(-17) \\
1.36(+0) & 6.13(-2) & 7.27(-17) & -7.63(-18) & +6.92(-18) & -8.43(-18) &
+8.40(-18) \\
1.42(+0) & 6.61(-2) & 8.52(-17) & -5.89(-18) & +5.11(-18) & -9.50(-18) &
+9.45(-18) \\
1.49(+0) & 7.12(-2) & 5.52(-17) & -7.17(-18) & +7.35(-18) & -6.92(-18) &
+6.93(-18) \\
1.57(+0) & 7.59(-2) & 6.39(-17) & -4.96(-18) & +5.08(-18) & -7.61(-18) &
+7.62(-18) \\
1.64(+0) & 7.99(-2) & 5.35(-17) & -6.73(-18) & +6.77(-18) & -6.75(-18) &
+6.78(-18) \\
1.73(+0) & 8.36(-2) & 4.50(-17) & -4.13(-18) & +4.14(-18) & -6.05(-18) &
+6.06(-18) \\
1.81(+0) & 8.75(-2) & 3.97(-17) & -4.34(-18) & +4.14(-18) & -5.68(-18) &
+5.66(-18) \\
1.90(+0) & 9.16(-2) & 3.66(-17) & -3.80(-18) & +4.03(-18) & -5.24(-18) &
+5.25(-18) \\
2.00(+0) & 9.58(-2) & 4.98(-17) & -5.83(-18) & +5.97(-18) & -6.40(-18) &
+6.43(-18) \\
2.09(+0) & 1.00(-1) & 3.36(-17) & -3.95(-18) & +3.63(-18) & -4.96(-18) &
+4.96(-18) \\
2.20(+0) & 1.05(-1) & 3.45(-17) & -3.79(-18) & +4.12(-18) & -4.98(-18) &
+4.98(-18) \\
2.30(+0) & 1.09(-1) & 3.32(-17) & -4.13(-18) & +3.58(-18) & -4.86(-18) &
+4.85(-18) \\
2.41(+0) & 1.14(-1) & 2.87(-17) & -3.42(-18) & +3.34(-18) & -4.45(-18) &
+4.45(-18) \\
2.53(+0) & 1.19(-1) & 3.93(-17) & -3.74(-18) & +3.71(-18) & -5.23(-18) &
+5.23(-18) \\
2.65(+0) & 1.24(-1) & 3.66(-17) & -3.91(-18) & +4.07(-18) & -5.01(-18) &
+5.02(-18) \\
2.78(+0) & 1.30(-1) & 3.06(-17) & -4.08(-18) & +3.55(-18) & -4.60(-18) &
+4.59(-18) \\
2.91(+0) & 1.35(-1) & 1.68(-17) & -2.59(-18) & +2.26(-18) & -3.62(-18) &
+3.61(-18) \\
3.05(+0) & 1.41(-1) & 1.09(-17) & -3.02(-18) & +3.28(-18) & -3.42(-18) &
+3.44(-18) \\
3.19(+0) & 1.47(-1) & 4.94(-18) & -2.52(-18) & +2.99(-18) & -2.75(-18) &
+2.81(-18) \\
3.34(+0) & 1.53(-1) & 4.00(-18) & -2.58(-18) & +3.09(-18) & -2.49(-18) &
+2.49(-18) \\
3.50(+0) & 1.59(-1) & 4.18(-18) & -2.06(-18) & +2.37(-18) & -2.36(-18) &
+2.37(-18) \\
3.66(+0) & 1.65(-1) & 9.06(-18) & -2.80(-18) & +2.95(-18) & -3.04(-18) &
+3.05(-18) \\
3.83(+0) & 1.72(-1) & 3.31(-18) & -1.81(-18) & +2.36(-18) & -2.03(-18) &
+2.10(-18) \\
4.01(+0) & 1.79(-1) & 5.15(-18) & -2.49(-18) & +2.87(-18) & -2.30(-18) &
+2.35(-18) \\
4.19(+0) & 1.85(-1) & 8.29(-18) & -2.32(-18) & +2.01(-18) & -2.59(-18) &
+2.59(-18) \\
4.38(+0) & 1.92(-1) & 8.17(-18) & -2.42(-18) & +2.59(-18) & -2.49(-18) &
+2.49(-18) \\
4.57(+0) & 1.99(-1) & 9.22(-18) & -2.15(-18) & +2.14(-18) & -2.50(-18) &
+2.51(-18) \\
4.78(+0) & 2.07(-1) & 1.57(-17) & -2.37(-18) & +2.65(-18) & -3.04(-18) &
+3.09(-18) \\
4.99(+0) & 2.14(-1) & 1.13(-17) & -1.88(-18) & +2.11(-18) & -2.59(-18) &
+2.62(-18) \\
5.20(+0) & 2.21(-1) & 1.32(-17) & -2.15(-18) & +1.97(-18) & -2.77(-18) &
+2.73(-18) \\
5.43(+0) & 2.29(-1) & 1.64(-17) & -1.49(-18) & +1.52(-18) & -2.90(-18) &
+2.90(-18) \\
5.66(+0) & 2.36(-1) & 1.56(-17) & -1.90(-18) & +1.96(-18) & -2.77(-18) &
+2.77(-18) \\
5.90(+0) & 2.44(-1) & 1.95(-17) & -1.43(-18) & +1.68(-18) & -3.05(-18) &
+3.07(-18) \\
6.15(+0) & 2.52(-1) & 1.85(-17) & -1.49(-18) & +1.43(-18) & -2.93(-18) &
+2.93(-18) \\
6.40(+0) & 2.59(-1) & 2.50(-17) & -1.53(-18) & +1.69(-18) & -3.41(-18) &
+3.42(-18) \\
6.67(+0) & 2.67(-1) & 2.47(-17) & -2.38(-18) & +2.33(-18) & -3.38(-18) &
+3.39(-18) \\
6.94(+0) & 2.75(-1) & 2.59(-17) & -2.72(-18) & +2.48(-18) & -3.42(-18) &
+3.41(-18) \\
7.22(+0) & 2.85(-1) & 2.76(-17) & -2.66(-18) & +2.34(-18) & -3.54(-18) &
+3.53(-18) \\
7.51(+0) & 2.97(-1) & 2.39(-17) & -2.32(-18) & +2.85(-18) & -3.19(-18) &
+3.23(-18) \\
7.81(+0) & 3.11(-1) & 2.45(-17) & -2.06(-18) & +2.18(-18) & -3.20(-18) &
+3.21(-18) \\
8.13(+0) & 3.26(-1) & 3.27(-17) & -2.54(-18) & +2.38(-18) & -3.86(-18) &
+3.84(-18) \\
8.47(+0) & 3.42(-1) & 2.96(-17) & -2.36(-18) & +2.66(-18) & -3.55(-18) &
+3.56(-18) \\
8.82(+0) & 3.59(-1) & 3.70(-17) & -2.27(-18) & +2.55(-18) & -4.15(-18) &
+4.18(-18) \\
9.18(+0) & 3.76(-1) & 4.29(-17) & -2.34(-18) & +2.20(-18) & -4.63(-18) &
+4.63(-18) \\
9.57(+0) & 3.95(-1) & 3.75(-17) & -2.88(-18) & +2.49(-18) & -4.22(-18) &
+4.18(-18) \\
9.97(+0) & 4.16(-1) & 4.41(-17) & -2.36(-18) & +2.38(-18) & -4.72(-18) &
+4.72(-18) \\
1.04(+1) & 4.37(-1) & 3.71(-17) & -2.04(-18) & +2.33(-18) & -4.13(-18) &
+4.14(-18) \\
1.08(+1) & 4.60(-1) & 3.76(-17) & -2.03(-18) & +1.80(-18) & -4.17(-18) &
+4.17(-18) \\
1.13(+1) & 4.85(-1) & 3.50(-17) & -2.27(-18) & +2.12(-18) & -3.96(-18) &
+3.96(-18) \\
1.18(+1) & 5.11(-1) & 3.18(-17) & -1.80(-18) & +1.77(-18) & -3.71(-18) &
+3.71(-18) \\
1.25(+1) & 8.11(-1) & 3.21(-17) & -1.59(-18) & +1.94(-18) & -3.75(-18) &
+3.75(-18) \\
\enddata
\tablecomments{\doublespace
The binned cross section is given as a function of bin energy center $E_{\rm
center}$. Also listed is the energy width $E_{\rm width}$ of each bin. The bin
start and end energies can be obtained as $E_{\rm center}\mp0.5E_{\rm width}$.
The asymmetric statistical and systematic uncertainties are given by $U_{\rm
stat}$ and $U_{\rm syst}$, respectively. The superscripts ``lo'' and ``hi'' give
the lower and upper limits, respectively, for the corresponding uncertainties.
The total systematic error in the table includes the 8.4\%  absolute scaling
error and the uncertainties from background correction, detector geometrical
efficiency correction, and $\ktperp^{\rm T}$ and $\ktpar^{\rm T}$ (see text).
The format $x(y)$ signifies $x\e{y}$.
}
\end{deluxetable}

\begin{deluxetable}{crll}
\tablecolumns{3} 
\tablewidth{0pc} 
\tablecaption{ Fit parameters for the \NHp DR plasma rate coefficient
$\alpha_{\rm pl}$ using Equation~(\ref{eq:plasmafitnew}). 
		\label{tab:plasmares}
		}
\tablehead{ \colhead{Parameter} & \multicolumn{2}{c}{Value}  &
\colhead{Units}\\
\cline{2-3}
 & \colhead{$x$}& \colhead{$y$} & 
}
\startdata
$A$	& \phs$2.11$	& $-7$ \phn & ${\rm cm^3\ s^{-1}}$ \\
$n$	& \phs$7.90$	& $-1$ & dimensionless\\
$c_1$	& $-1.12$	& $-4$ & ${\rm K^{3/2}\ cm^3\ s^{-1}}$ \\
$c_2$	& $-2.49$	& $-4$ & ${\rm K^{3/2}\ cm^3\ s^{-1}}$ \\
$c_3$	& $-9.14$	& $-4$ & ${\rm K^{3/2}\ cm^3\ s^{-1}}$ \\
$c_4$	& \phs$8.05$& $-3$ & ${\rm K^{3/2}\ cm^3\ s^{-1}}$ \\
$T_1$	& \phs$1.28$	& \phm{$-$}1& K \\
$T_2$	& \phs$9.24$	& \phm{$-$}1& K \\
$T_3$	& \phs$4.81$	& \phm{$-$}2& K \\
$T_4$	& \phs$5.03$	& \phm{$-$}3& K \\
\enddata
\tablecomments{\doublespace
The value for each parameter is given by $x\e{y}$. }
\end{deluxetable}

\clearpage


\clearpage
\clearpage

\clearpage

\end{document}